\documentclass[12pt,a4paper,final]{iopart}
\usepackage{iopams}  
\usepackage{graphicx}
\usepackage[breaklinks=true,colorlinks=true,linkcolor=blue,urlcolor=blue,citecolor=blue]{hyperref}
\usepackage{float}
\usepackage{subcaption}
\usepackage{lipsum}
\usepackage{notoccite}
\usepackage{amsmath}
\usepackage{xcolor}
\usepackage[normalem]{ulem}
\usepackage{multirow}
\usepackage{cleveref}

\begin{document}

\title[Retrograde Polish Doughnuts around Boson Stars]{Retrograde Polish Doughnuts around Boson Stars}

\author{Matheus C. Teodoro}
\address{Institute of Physics, University of Oldenburg, 26111 Oldenburg, Germany}
\ead{matheus.do.carmo.teodoro@uol.de}

\author{Lucas G. Collodel}
\address{Theoretical Astrophysics, University of T\"ubingen, 72076 T\"ubingen, Germany}
\ead{lucas.gardai-collodel@uni-tuebingen.de}

\author{Jutta Kunz}
\address{Institute of Physics, University of Oldenburg, 26111 Oldenburg, Germany}
\ead{jutta.kunz@uni-oldenburg.de}


\begin{abstract}
We investigate polish doughnuts with a uniform constant specific angular momentum distribution in the space-times of rotating boson stars. In such space-times thick tori can exhibit unique features not present in Kerr space-times. For instance, in the context of retrograde tori, they may possess two centers connected or not by a cusp. Rotating boson stars also feature a static ring, neither present in Kerr space-times. This static ring consists of static orbits, where particles are at rest with respect to a zero angular momentum observer at infinity. Here we show that the presence of a static ring allows for an associated static surface of a retrograde thick torus, where inside the static surface the fluid moves in prograde direction. We classify the retrograde Polish doughnuts and present several specific examples.
\end{abstract}

\section{Introduction}

The topic of accretion disks is manifold. From a diverse set of analytical models to general relativistic magnetohydrodynamics (GRMHD) simulations, the study of accretion disks is deeply related to a variety of astrophysical scenarios. In fact, by transforming gravitational potential energy into radiation, accretion disks provide insight for the emission of X-ray binaries, active galactic nuclei (AGN) and quasars as well as input on the search for evidence of black holes (BHs) (some review papers can be found at \cite{Narayan:2013gca,Abramowicz:2011xu}). The efforts done to study these disks have recently played a crucial role, for instance, in the seminal imaging of the compact object at the center of M87 \cite{Akiyama:2019cqa,Akiyama:2019brx,Akiyama:2019sww,Akiyama:2019bqs,Akiyama:2019fyp,Akiyama:2019eap}, providing simulations of disk models around Kerr BHs later compared with the data acquired by the Event Horizon Telescope Collaboration (EHTC). Furthermore, the imaging of the compact object at the center of M87 brings into clear prospect to achieve the imaging of Sgr A* \cite{Mizuno:2018lxz}, as well.

Albeit Kerr BHs are largely assumed to be the most astrophysically relevant kind of compact objects to describe the supermassive entities at the center of galaxies, alternative contenders must be considered in order to find where and how deviations from observational signatures might arise. Within the realms of general relativity (GR), boson stars (BSs) are possibly the most promising candidates to BH mimickers \cite{Schunck:1997dn,Schunck:2008xz,Schunck:1999pm,Torres:2000dw,Mielke:2000mh,Guzman:2005as,Guzman:2005bs,Vincent:2015xta,Olivares:2018abq,Vincent:2020dij}. The mass of these exotic objects has bounds determined by the nature of their self-interaction and the mass of the constituent scalar field, and overall can span a scale that goes from atomic order to that of supermassive BHs dwelling in the center of galaxies
(see e.g. the reviews on BSs \cite{Jetzer:1991jr,Schunck:2003kk,Liebling:2012fv}).

BSs feature very peculiar properties. Typical among these is the lack of a hard surface, thus there is no specific envelope where the pressure becomes zero. Consequently, these objects extend to spatial infinity,
although with an exponentially decaying scalar field.
Accordingly there is no exterior vacuum space-time surrounding them. The building block of BSs - the massive complex scalar field - interacts only gravitationally with ordinary matter, which can therefore freely orbit their interior. Since a well-defined surface is missing, the usual definition of compactness as the ratio of mass to radius, $\mathcal{C}=M/R$, is no longer unique for BSs, but depends on the chosen definition of the BS radius.
Nevertheless, independent of the chosen definition, many BSs qualify as compact objects (see, e.g., \cite{Kleihaus:2011sx}), with some BSs featuring even light-rings, and giving rise to multiple images of the celestial sphere \cite{Cunha:2015yba}.

Inspired by Wheeler's \emph{geon} \cite{Wheeler:1955zz}, BSs were first obtained half a century ago \cite{PhysRev.172.1331,PhysRev.168.1445,PhysRev.187.1767}, and many of their aspects have been thereon extensively studied, as, e.g., spinning BS configurations \cite{PhysRevD.50.7721,PhysRevD.56.762,Schunck:1996he,PhysRevD.72.064002,PhysRevD.77.064025,Collodel:2017biu,Collodel:2019ohy}, stability and formation \cite{Lee:1988av,Kusmartsev:1990cr,Kleihaus:2011sx,Sanchis-Gual:2019ljs}, or geodesic motion around them \cite{Eilers:2013lla,Brihaye:2014gua,Grandclement:2014msa,Meliani:2015zta,Grould:2017rzz,Collodel:2017end}.
Boson stars could form through a process called gravitational cooling, where an initial cloud of scalar matter undergoes gravitational collapse and ejects matter and angular momentum through a series of cascade events until it settles into an equilibrium configuration. Conversely, if this state is unstable, the system would either continue collapsing until a black hole is formed, or have all of its contents dispersed away.
Although BSs do not possess a shadow in the sense of  its classical definition, they nevertheless feature a dark region very similar to the shadow of a black hole, when considered in an analogous astrophysical environment, with the light emission dominantly sourced by an accretion disk \cite{Vincent:2015xta,Olivares:2018abq,Vincent:2020dij}. 
Thence, the imaging of a BS might be very similar to that of a BH and different types of observations might be necessary to tell one from the other, as many parameters must be constrained such as mass, angular momentum, scalar charge and self-interaction potential. In \cite{Olivares:2018abq}, the authors find that the dark region resembling the shadow of a BH in the image due to a non-rotating BS is too small to fit the data from M87, but further investigation is required before dismissing the possibility of it being a rotating BS.
Regarding fluid configurations around BSs, not only has research been done in the direction of disks \cite{Meliani:2015zta,Vincent:2015xta,Olivares:2018abq,Vincent:2020dij} but also on tidal disruption events and disk formation \cite{Meliani:2017ktw,Teodoro:2020gps}.

Although diverse, much of the research done towards accretion disks tends to have some common grounds. An example are the so called ``Polish doughnuts". These thick tori represent one of the simplest analytical fluid configurations in non-geodesic motion around compact objects. Polish doughnuts were initially built considering isentropic perfect fluids \cite{1976ApJ...207..962F} and then generalized for barotropic perfect fluids \cite{1978A&A....63..221A,1978A&A....63..209K}.
Both in the analytical and numerical contexts, these equilibrium non-self-gravitating fluid configurations are commonly used as a first step from which robuster analysis can arise. In the context of GRMHD simulations, for example, they are used as initial condition to investigate advection-dominated flows \cite{Narayan:2012yp}, weakly magnetized disks \cite{2004ApJ...611..977M}, jet formation \cite{2006MNRAS.368.1561M}, similarities and differences between BH and BS images \cite{Vincent:2015xta,Olivares:2018abq,Vincent:2020dij} and so on. The doughnuts are also a starting point for many analytical studies, among which we would like to highlight the ones regarding accretion disks sheltered by less-mainstream spacetimes, like  Kerr-de Sitter backgrounds \cite{Slany:2005vd}, Kerr black holes with scalar hair \cite{2019PhRvD..99d3002G},  distorted static BHs \cite{Faraji:2020xzo}, deformed compact objects \cite{Faraji:2020tmv} and also BSs \cite{Meliani:2015zta}.

Interestingly, when surrounding BSs these tori can exhibit unique features, such as not having a cusp or being endowed with two centers \cite{Meliani:2015zta}. In this context, and motivated by the fact that BSs can possess a \emph{static ring} \cite{Collodel:2017end}, i.e., a set of orbits that are at rest with respect to a zero angular momentum observer (ZAMO) at infinity, we here present a first approach to retrograde Polish Doughnuts around BSs, for which \emph{static surfaces} can be observed. These static surfaces enclose a region, where the fluid moves in prograde direction, whereas outside the static surfaces the fluid is in retrograde motion.
 
In this paper we consider retrograde geometrically thick tori around various models of rotating BSs. After recalling some properties of rotating BSs in section 2, including their static ring, we discuss the building of thick torus solutions in section 3. Next we address the Keplerian and rest specific angular momenta in section 4 for the set of rotating BS spacetimes and discuss their influence on the building of torus solutions. In section 5 we present our classification scheme of thick retrograde tori and give several instructive examples. Section 6 contains our conclusion.







\section{Rotating boson stars}
\label{sec:rbs}

Here we consider BSs obtained in the Einstein-Klein-Gordon model described by the action
\begin{equation}
  \label{action}
S = \int{\sqrt{-g}\left(\frac{R}{16 \pi G}-\mathcal{L}_\mathrm{m}\right) d^4 x},  
\end{equation}
where $R$ is the curvature scalar with respect to the metric $g_{\mu\nu}$, $g$ is the metric determinant, $G$ is Newton's constant, and 
$\mathcal{L}_\mathrm{m}$ is the Lagrangian of the complex boson field $\phi$
\begin{equation}
  \label{lagFLS}
\mathcal{L}_\mathrm{m} = \left|\partial_\mu\phi\right|^2 + m^2|\phi|^2,  
\end{equation}
with scalar mass $m$. We do not take any self-interaction of the scalar field into account.

The action (\ref{action}) has a  global $\mathrm{U}(1)$ invariance associated with the complex scalar field, $\phi\to\phi e^{i\chi }$, with constant $\chi$. This symmetry gives rise to the
conserved Noether current 
\begin{equation}
    j_\mu = i(\phi\partial_\mu\phi^\ast-\phi^\ast\partial_\mu\phi) ,
  \label{Noether}  
\end{equation}
with conserved charge $Q=\int{\sqrt{-g}j^t d^3 x}$.

Variation of the action (\ref{action}) with respect to the metric 
and the scalar field leads to the coupled set of Einstein-Klein-Gordon equations,
\begin{equation}
    \label{Einstein}
R_{\mu\nu}-\frac{1}{2}R g_{\mu\nu}={8\pi G}\, T_{\mu\nu},
\end{equation}
with stress-energy tensor
\begin{equation}
    \label{SET}
T_{\mu\nu}=\left(\partial_{\mu}\phi\partial_{\nu}\phi^\ast
+\partial_{\nu}\phi\partial_{\mu}\phi^\ast\right) -\mathcal{L}_\mathrm{m}g_{\mu\nu} ,
\end{equation}
and
\begin{equation}
    \label{scaleq}
    \left(\Box -m^2\right)\phi=0 ,
\end{equation}
with covariant d'Alembert operator $\Box$.

Stationary spinning axially-symmetric BS space-times possess two commuting Killing vectors associated with the time and azimuthal coordinates $t$ and $\varphi$, respectively. In these coordinates the metric can be written in the form
\begin{equation}
    ds^2 = -\alpha^2 dt^2
    + A^2 \left( dr^2 + r^2 d\theta^2 \right)
    + B^2 r^2 \sin^2 \theta \left( d\varphi + \beta^\varphi dt \right)^2 ,
    \label{metric}
\end{equation}
where $\alpha$ is the lapse function, $\beta^\varphi$ is the shift function, and the four functions $A$, $B$, $\alpha$, $\beta$ depend on $r$ and $\theta$, only.
The scalar field, however, must depend of all four spacetime coordinates. From eq.~(\ref{Noether}) it follows that if $\phi$ is independent of time, then the conserved charge is zero: $\phi\neq\phi(t)\Longrightarrow Q=0$, and the system would be inherently unstable. Furthermore, it follows from eq.~(\ref{SET}) that the scalar field must depend on the azimuthal angle $\varphi$ in order to rotate and source the dragging of the spacetime. The only way this field can depend on $t$ and $\varphi$ without jeopardizing the stationarity and axisymmetry of the spacetime is harmonically, hence 
\begin{equation}
   \label{scalans}
\phi=f(r,\theta)e^{i(\omega t-k\varphi)}, 
\end{equation}
where the constant $\omega$ is the angular frequency of the scalar field, and 
$k$ the azimuthal winding number which must be an integer due to the identification $\phi(\varphi=0)=\phi(\varphi=2\pi)$. Solutions only exist for a bounded set of frequency values, which poses an eigenvalue problem for the system. The winding number tells how many nodes the real and the imaginary parts of the scalar field have  along the $\vec{\partial}_\varphi$ direction, and hence how strong its angular excitation is, which has an impact on the total angular momentum of the BS, as seen below.  

Substitution of the ansatz into the field equation then leads to the coupled set of equations for the functions to be solved numerically with the professional FIDISOL/CADSOL package \cite{schoen}, based on a Newton-Raphson scheme, subject to an appropriate set of boundary conditions. The set of PDEs is presented in explicit form according to the parametrization eq.~(\ref{metric}) and (\ref{scalans}) in the \ref{app:fe}.
The metric functions approach asymptotically Minkowski space-time, while the scalar field vanishes exponentially,
being proportional to $e^{-\sqrt{m^2-\omega^2}r}/r$, 
for $\omega < m$.

Axial symmetry and regularity impose boundary conditions on the symmetry axis, $\theta=0,\pi$.
Since the solutions are symmetric with respect to reflections in the equatorial plane the range of values
of the angular variable $\theta$ can be restricted to $\theta \in [0,\pi/2]$ accompanied by an appropriate choice of boundary conditions in the equatorial plane.

Asymptotic expansions of the metric functions at spatial infinity yield the total mass $M$ and angular momentum $J$ of the rotating BSs
\begin{equation}
    \label{ADM}
g_{tt}=-1 + \frac{2 G M}{ r}+O\left(\frac{1}{r^2}\right), \quad
g_{t\varphi}=-\frac{2 G J}{r}\sin^2 \theta+O\left(\frac{1}{r^2}\right).
\end{equation}
Note that there is a quantization relation for the angular momentum $J$ of the BSs,
$J=k Q$, with Noether charge $Q$ 
and winding number $k$.

\begin{figure}[t]
 \hspace{-1.5cm}\includegraphics[scale=0.5]{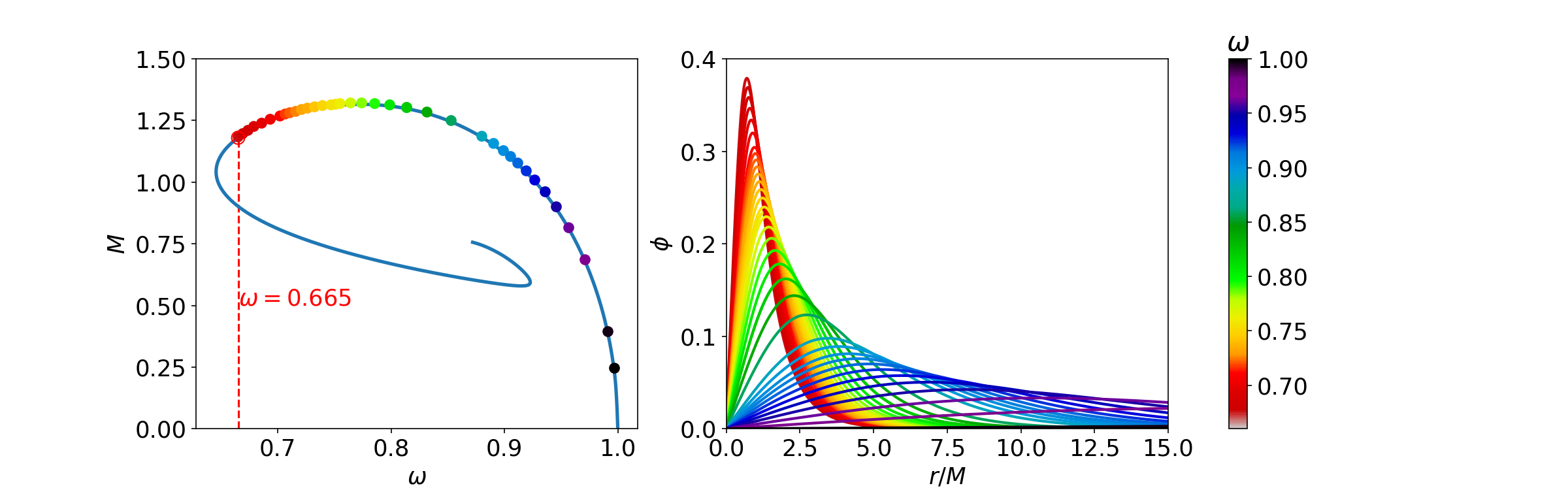}%
     \caption{BS properties ($k=1)$: mass $M$ vs frequency $\omega$ (a) and modulus of the scalar field $\phi$ in the equatorial plane vs scaled radial coordinate $r/M$ (b), color coded according to $\omega$. BSs beyond $\omega=0.665$ possess ergoregions.}
     \label{fig:solutions}
\end{figure}

We will now restrict to the set of (parity even) rotating BSs with the lowest winding number $k=1$.
In Fig.~\ref{fig:solutions}(a) we exhibit their mass $M$ versus their angular frequency $\omega$. The fundamental branch extends from the vacuum at $\omega=m$ to the minimal frequency $\omega_\text{min}$, where it bifurcates with the second branch. The system of equations (\ref{Einstein})-(\ref{scaleq}) features a scaling symmetry with the boson mass, which we make use of by setting $r\rightarrow rm$, $\omega\rightarrow\omega/m$, from which it follows that the total mass scales as $M\rightarrow Mm$.
(From now on we will employ the scaled quantities. Please note that $M$ has units of $M_P^2/m$, where $M_P$ is the Planck mass.)
The circle at $\omega=0.665$ denotes the onset of ergo-regions, which are present for BSs beyond this point. Here we focus on BSs without ergo-regions and thus BSs along the fundamental branch with $\omega>0.665$. 
As aforementioned, the higher the winding number is, the more excited the field gets, and consequently its total mass and angular momentum for a particular choice of $\omega$ or $Q$ become larger. This is depicted in Fig.~6 of \cite{Grandclement:2014msa} which we refer the reader to for a comparison of the global charges for different values of $k$.

We show the modulus of the scalar field $\phi$ in the equatorial plane versus the scaled radial coordinate $r/M$ in Fig.~\ref{fig:solutions}(b), color coded according to the frequency $\omega$ in the considered range $0.665 < \omega < 1$. With decreasing $\omega$ the maximum of the scalar field modulus increases considerably and shifts to smaller values of $r/M$. Thus the tori formed by the scalar field become stronger localized while moving closer to the center. Note, that we are employing a quasi-isotropic radial coordinate here.

\begin{figure}[h]
\centering
\begin{subfigure}{.5\textwidth}
  \centering
  \includegraphics[width=1\linewidth]{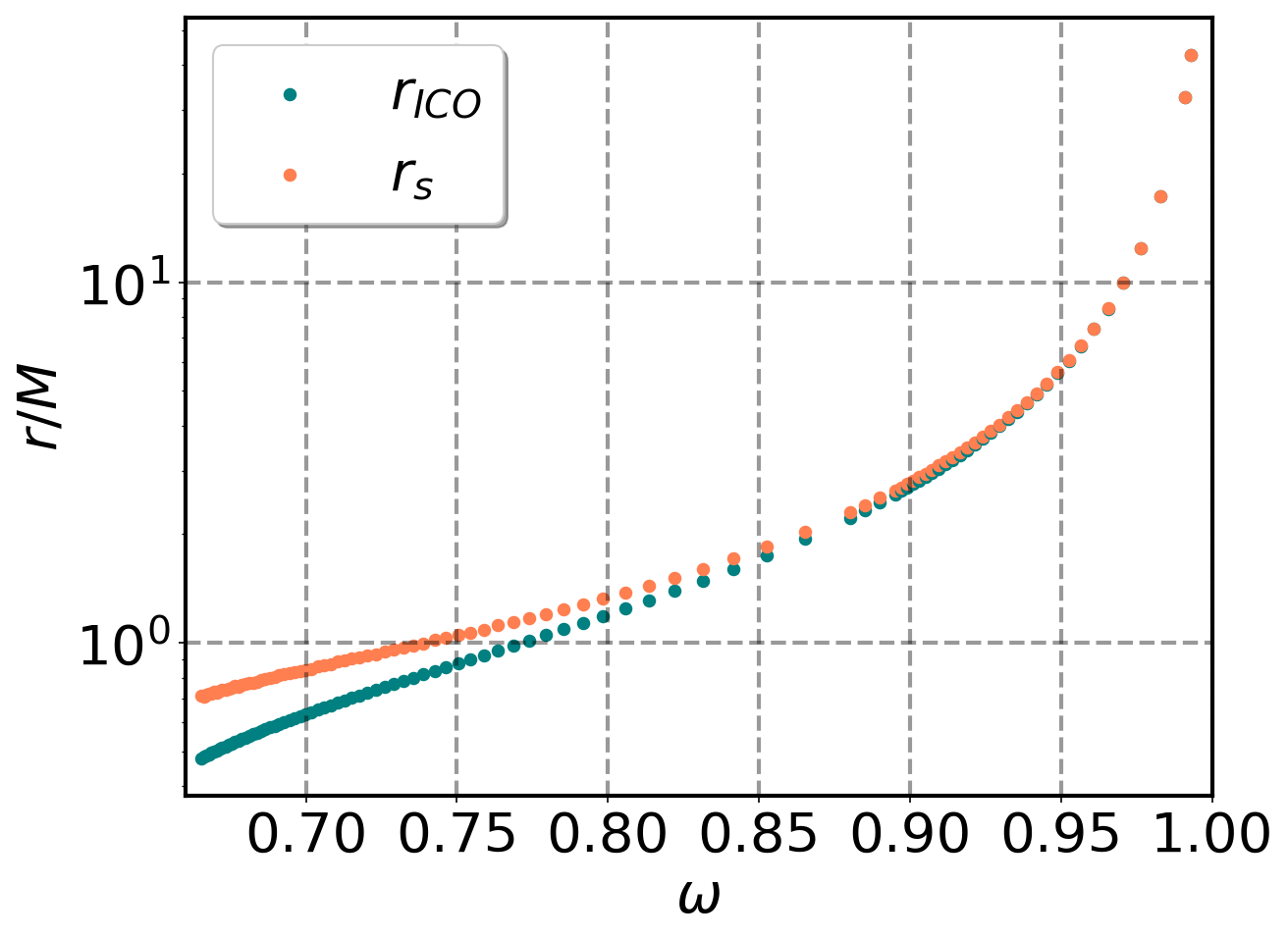}
  \caption{ }
       \label{fig:srbs}
\end{subfigure}%
\begin{subfigure}{.5\textwidth}
  \centering
  \includegraphics[width=1\linewidth]{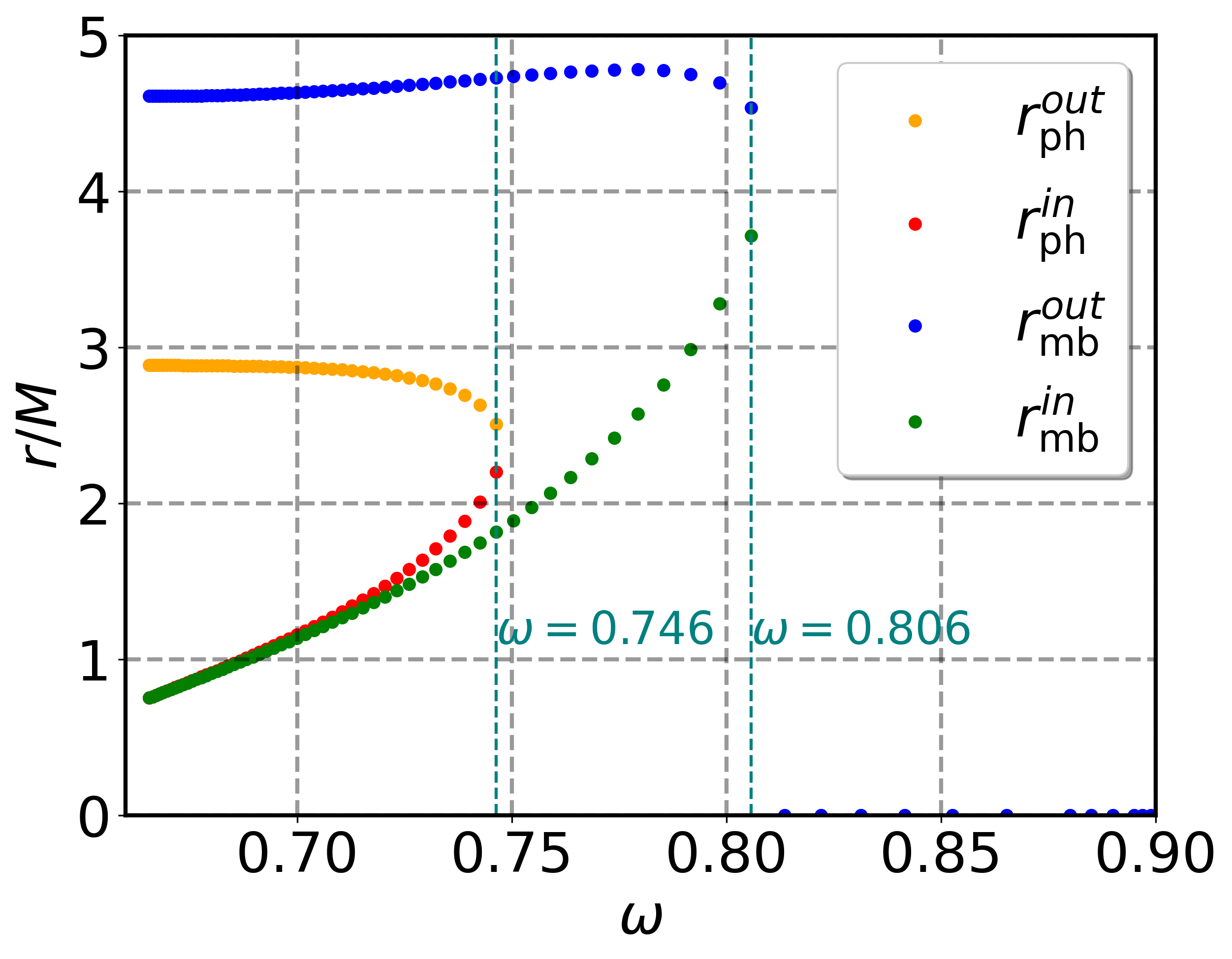}
 \caption{ }
       \label{fig:photonorbits}
\end{subfigure}
     \caption{(a) Normalized position $r_{s}$ of the static ring vs frequency $\omega$ for the rotating BSs ($k=1$) without ergoregions (orange) and normalized position $r_{\rm ICO}$ of the innermost circular orbit (see section 3) for the same set of solutions (cyan).
     (b) Normalized location of the innermost (red) and outermost (yellow) photon orbits, when existing ($\omega\le 0.746$), and normalized innermost (green) and outermost (blue) marginally stable orbits, when existing ($\omega\le 0.806$).}
     \label{fig:ICOsandPO}
\end{figure}

Due to their off-center field (and consequently energy density) distribution, rotating boson stars warp the space-time around them in such peculiar manner that, although not exclusive to these objects, it causes particle motion in this background to be somewhat non-intuitive. Particularly, all the solutions presented in the figures above feature a \emph{static ring}, i.e., a ring of points in the equatorial plane where a particle at rest with respect to a ZAMO at infinity remains at rest in a geodesic motion described by a circular orbit with zero angular velocity \cite{Collodel:2017end}. Moreover, these orbits are semi-stable, and if the particle of interest is perturbed, albeit not returning to rest, it engages in a slow movement oscillating around the static ring while orbiting the central object. The necessary and sufficient conditions for the appearance of a static ring are stationarity, axisymmetry, circularity, and furthermore the $g_{tt}$ component of the metric must contain a local maximum (with our metric signature) off center where $g_{tt}<0$, since we are concerned about timelike geodesics. In Fig.~\ref{fig:srbs} we display the position of the static ring, normalized by the star's mass, with respect to the frequency for the set of solutions containing no ergoregions. The position of the static ring decreases monotonically with $\omega$, as the star's compactness increases.

\section{Building thick torus solutions}

In this section we present the general recipe for constructing non-self-gravitating thick tori with a constant angular momentum distribution. Having its roots in Boyer's first approach to the problem \cite{boyer_1965}, the solutions we shall present were developed first for the Schwarzschild metric in \cite{1978A&A....63..221A}  and later for the Kerr metric in 
\cite{1978A&A....63..209K}. These solutions are minimalist in the sense that they consider nothing but the gravity of a central object and a perfect fluid rotating around it. 

These thick torus solutions arise by assuming an axisymmetric stationary fluid distribution in an also axisymmetric stationary space-time, for which the metric tensor reads,
\begin{align}
    ds^2=g_{tt}dt^2+2g_{t\varphi}dtd\varphi+g_{rr}dr^2+g_{\theta\theta}d\theta^2+g_{\varphi\varphi}d\varphi^2.
    \label{eq:generalmetric}
\end{align}
When the space-time is endowed with the Killing vectors $\eta^\mu=\delta^\mu_t$ and $\xi^\mu=\delta^\mu_\varphi$, the fluid four-velocity can be written as,
\begin{equation}
u^\mu=u^t(\eta^\mu+\Omega \xi^\mu)
\end{equation}
where the angular velocity is defined by $\Omega\dot{=}\frac{u^\varphi}{u^t}$.
Requiring $u^\mu$ to be normalized and applying the properties of the Killing vectors the fluid four-acceleration can be written as
\begin{equation}
a_\mu=\partial_\mu|\ln(u_t)|-\frac{\Omega}{1-\Omega l}\partial_\mu l ,
\label{ac}
\end{equation}
where $l\dot{=}-\frac{u_\varphi}{u_t}$ is the fluid specific angular momentum. Although different definitions of $l$ were initially considered in the early development of this type of torus \cite{1976ApJ...207..962F} a set of motivations for adopting this definition can be found in \cite{1978A&A....63..209K}.

Considering non-self-gravitating fluids with a barotropic equation of state with stress-energy tensor
\begin{align}
    T_{\mu\nu}=\rho h u_\mu u_\nu +pg_{\mu\nu},
\end{align}
where $p$, $h$ and $\rho$ are the fluid pressure, specific enthalpy and rest-mass density, the fluid Euler equations read
\begin{equation}
-\frac{1}{\rho h}\partial_\mu p=\partial_\mu|\ln(u_t)|-\frac{\Omega}{1-\Omega l}\partial_\mu l   .
\label{euler}
\end{equation}
We note that it is also possible to prove that for perfect barotropic fluids $\Omega=\Omega(r,\theta;l)$ \cite{1971AcA....21...81A,boyer_1965,Komissarov:2006nz}. This means that the surfaces of constant specific angular momentum and constant angular velocity coincide. Such surfaces are called {\it von Zeipel surfaces.}
In fact we can relate $\Omega$ and $l$ by
\begin{equation}
    \Omega=-\frac{g_{t\varphi}+g_{tt}l}{g_{\varphi\varphi}+g_{t\varphi}l},\,\,\,\,\,\,\,\,\,\,l=-\frac{g_{t\varphi}+g_{\varphi\varphi}\Omega}{g_{tt}+g_{t\varphi}\Omega}.
    \label{relation}
\end{equation}

Since $\Omega=\Omega(r,\theta;l)$, it is possible to integrate eq.~(\ref{euler}). It is useful to define an effective potential $\mathcal{W}(r,\theta)$ as
\begin{equation} 
    \mathcal{W}(r,\theta)-\mathcal{W}_{in}\dot{=}-\int_0^p \frac{dp'}{\rho h} \, ,
    \label{defW}
\end{equation}
where $\mathcal{W}_{in}$ is the potential calculated at the inner edge of the torus. Thus one finds
\begin{equation}
    \mathcal{W}(r,\theta)-\mathcal{W}_{in}=\ln{|u_t|}-\ln{|(u_t)_{in}|}-\int_{l_{in}}^l\frac{\Omega dl'}{1-\Omega l'} \, ,
\end{equation}
where $l_{in}$ is the specific angular momentum at the torus border. The integral on the right-hand side of the above equation vanishes by imposing $l(r,\theta)$ to be constant (hereafter $l_0$), producing {\it constant specific angular momentum thick torus} solutions, the subject of this paper. These solutions have the property of being marginally stable around Kerr BHs \cite{1975ApJ...197..745S} and commonly share the topological properties of other specific angular momentum distribution choices \cite{1980AcA....30....1J}.  

The choice of the specific angular momentum $l_0$ will then define completely the shape of the potential $\mathcal{W}$,
\begin{align}
    \mathcal{W}(r,\theta)=\ln\left(\frac{g_{t\varphi}^2-g_{tt}g_{\varphi\varphi}}{g_{\varphi\varphi}+2l_0g_{t\varphi}+l_0^2g_{tt}} \right)^{\frac{1}{2}} \, ,
    \label{potentialW}
\end{align}
while $\mathcal{W}_{in}$ is a constant that defines the outer border of the torus. The torus will then consist of layers of constant  $\mathcal{W}$, enveloped by the outermost equipotential defined by  $\mathcal{W}(r,\theta)=\mathcal{W}_{in}$.  Thus $\mathcal{W}_{in}<0$ represents finite tori,  while $\mathcal{W}_{in}>0$ represents infinite tori, i.e., these choices will lead to closed or open outermost equipotential surfaces, respectively. Physically feasible tori must be finite. This can be guaranteed by choosing the border of the fluid configuration to be closed, namely $\mathcal{W}_{in}<0$. The threshold condition $\mathcal{W}_{in}=0$ represents a torus closed at infinity, and for the sake of completeness we will present the torus solutions in section 5 with this choice.

The role of $\mathcal{W}$ can be better illustrated by considering an isentropic fluid with a polytropic equation of state, $p=\kappa\rho^{\Gamma},$ where $\kappa$ is the polytropic constant and $\Gamma$ the polytropic index. In that case one finds that $\frac{dp}{h\rho}=\frac{dh}{h}$, and thus from Eq.(\ref{defW}),
\begin{align}
    &h=\exp{(\mathcal{W}_{in}-\mathcal{W}(r,\theta))} ,\\
    &\rho(r,\theta)=\Big[\big(\frac{\Gamma-1}{\kappa\Gamma}\big)[\exp{(\mathcal{W}_{in} -\mathcal{W}(r,\theta))-1]\Big]^{1/(\Gamma-1)}} .
\end{align}
Consequently the equipotentials of $\mathcal{W}(r,\theta)$ will coincide with the constant rest density and constant pressure surfaces. Considering that the equatorial plane of the torus and the BS coincide, it is useful to analyze the behavior of $\mathcal{W}(r,\theta=\frac{\pi}{2})$ (hereafter $\mathcal{W}(r)$). 

The local minimum of $\mathcal{W}(r)$ corresponds to the local maximum of $\rho$ and $p$ and is usually referred to as \emph{the torus center}, while the local maximum of $\mathcal{W}(r)$ corresponds to the \emph{torus cusp}\footnote{Although slightly out of context for the solutions we shall present in this paper, we shall call a ``cusp" any point where an equipotential surface has a self-intersection.}. BH space-times commonly shelter torus solutions with one center and one cusp. In contrast, BSs can house torus configurations with more than one local potential minimum, being thus double-centered \cite{Meliani:2015zta}.  

Since the extrema of $\mathcal{W}(r)$ correspond to $\partial_r\mathcal{W}(r)=0$, in these locations $a_\mu=0$. Indeed, at these points of the equatorial plane the fluid experiences no pressure gradients and is thus moving geodesically. The corresponding geodesics must be circular since axisymmetry is assumed.  In this sense, addressing the Keplerian orbits around the gravitational source provides insight on the torus geometry. In particular the Keplerian specific angular momenta $l_{\text{K}}^{\pm}(r)$, where the $\pm$ signs correspond to the prograde and retrograde orbits, reveal the locations of the extrema of $\mathcal{W}(r)$, since these correspond to the radial coordinates where $l_{\text{K}}^{\pm}(r)=l_0$. 

In order to address the Keplerian orbits around the BSs of interest we have chosen to follow the approach introduced in \cite{Grandclement:2014msa} and later applied for a similar goal in \cite{Meliani:2015zta}. Following this approach the specific angular momenta for the circular orbits in the BS equatorial plane read
\begin{equation}
    l^\pm_{\rm K}=\frac{BrV_{\rm K}^\pm}{\alpha-\beta^\varphi BrV_{\rm K}^\pm},
    \label{lpm-Kepler}
\end{equation}
and the specific energy $E=-u_t$
\begin{align}
    E_{{\rm K}}^\pm=\frac{\alpha-\beta^\varphi BrV_{\rm K}^{\pm} }{\sqrt{1-V_{\rm K}^{\pm 2}}} ,
    \label{epm-Kepler}
\end{align}
where $B^2=\frac{g_{\varphi\varphi}}{r^2}$, $\beta^\varphi=\frac{g_{t\varphi}}{g_{\varphi\varphi}}$ and $\alpha^2=-g_{tt}+{\beta^\varphi}^2B^2r^2$. $V_{\rm K}^\pm$ is the orbit velocity with respect to the ZAMO
\begin{equation}
    V_{\rm K}^{\pm}=\frac{-\frac{Br}{\alpha}\partial_r\beta^\varphi\pm\sqrt{D}}{2\Big(\frac{\partial_rB}{B}+\frac{1}{r}\Big)},
    \label{zamoV}
\end{equation}
and the discriminant $D$ reads
\begin{equation}
    D=\frac{B^2r^2}{\alpha^2}(\partial_r\beta^\varphi)^2+4\partial_r(\ln{\alpha})\Big(\partial_r\ln{B}+\frac{1}{r}\Big).
    \label{discriminant}
\end{equation}

The Keplerian orbits exist only for $D\ge0$. Thus the \emph{Innermost Circular Orbit} (ICO) is defined by $D(r_{\rm ICO})=0$. The normalized position of the ICO for each BS solution can be found in Fig.~\ref{fig:srbs}.

Retrograde photon orbits can be found for some of the BSs here presented, namely the solutions for which $\omega\le 0.746$.  Given a BS, these orbital radii coincide with the locations where $V_{\rm K}^-(r)=-1$ and are depicted in Fig.~\ref{fig:photonorbits}. Furthermore, two marginally bound orbits can be found for solutions endowed with $\omega\le 0.806$. The location of such orbits correspond to the radial coordinate for which $E_{\rm K}^-(r)=1$ and can also be found in Fig.~\ref{fig:photonorbits}. No such features can be found in the prograde case.

For retrograde tori {\it static surfaces} are also feasible and will take part in the classification of the torus solutions. A static surface corresponds to the region of the torus in which the fluid stays at rest with respect to the ZAMO at infinity ($\Omega=0$).  
To find these surfaces it is useful to define the {\it rest specific angular momentum}, $l_r(r,\theta)$, by taking eq.~(\ref{relation}) and setting $\Omega=0$, leading to
\begin{equation}
l_r=-\frac{g_{t\varphi}}{g_{tt}}.
\label{restl}
\end{equation}
The static surfaces will be realized at the locations where $l_{r}(r,\theta)=l_0$. Since $l_{r}(r,\theta)$ assumes only negative values, such surfaces are more likely to exist for retrograde tori. 

\section{Keplerian and rest specific angular momenta}

\begin{figure}[h]
\centering
\begin{subfigure}{.5\textwidth}
  \centering
  \includegraphics[width=1\linewidth]{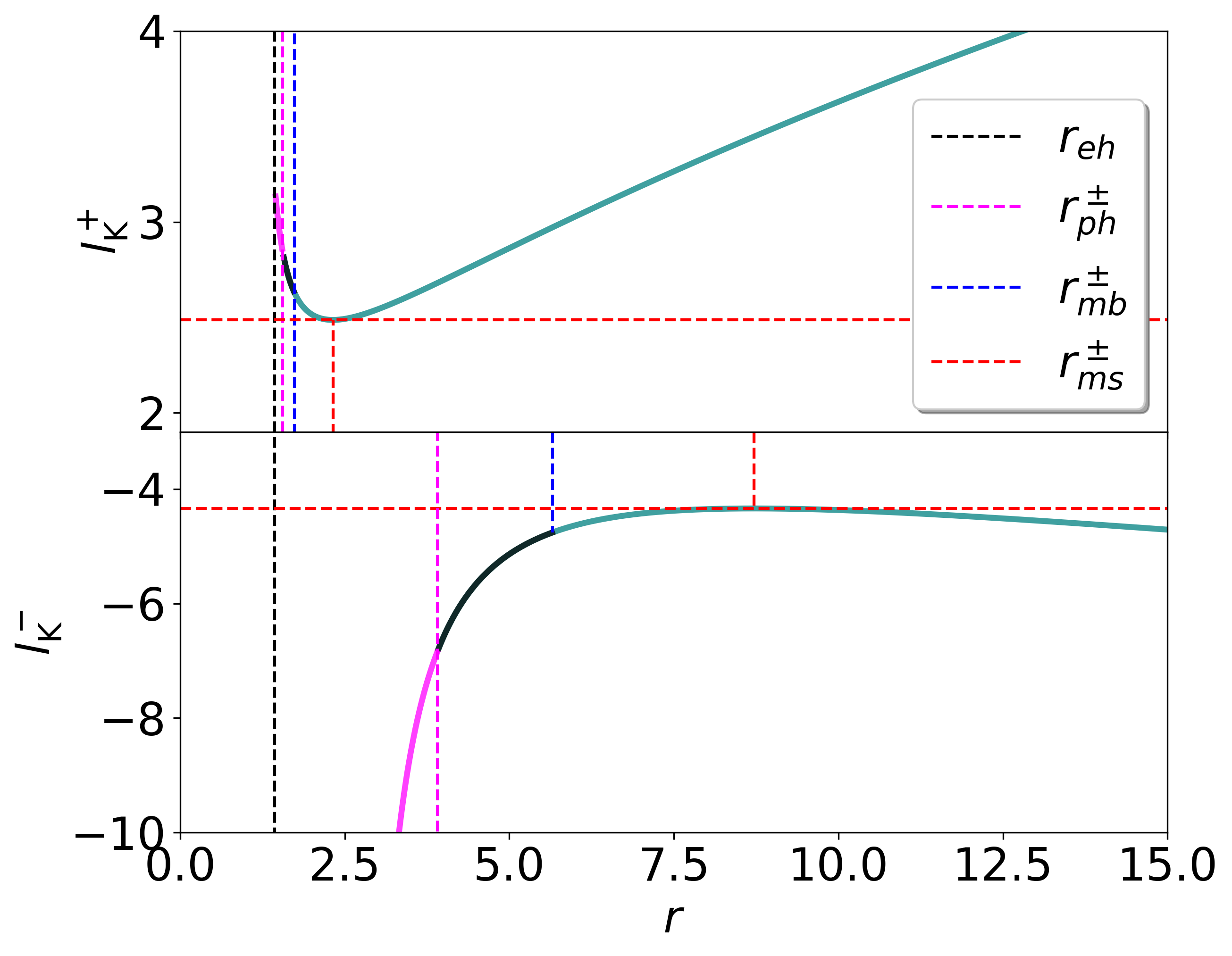}
  \caption{ }
  \label{fig:kerr}
\end{subfigure}%
\begin{subfigure}{.5\textwidth}
  \centering
  \includegraphics[width=1\linewidth]{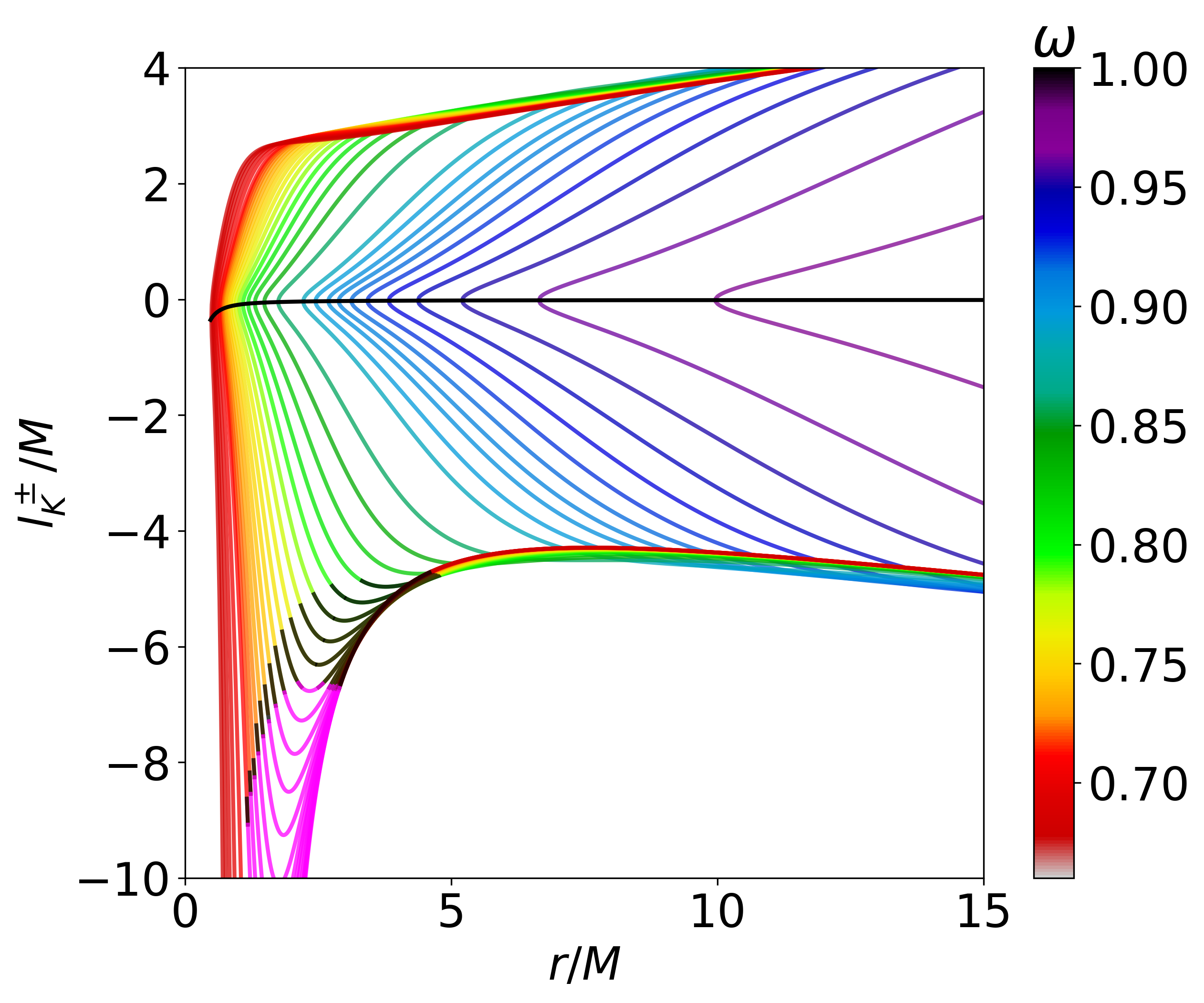}
 \caption{ }
  \label{fig:allK_BS}
\end{subfigure}
     \caption{(a) Specific angular momenta of Keplerian prograde (top) and retrograde (bottom) orbits for the Kerr BH with $a=0.9$ and $M=1$. Solid black lines represent unbound orbits, dotted lines represent the values for the marginally stable, marginally bound and photon orbits. The dotted black line represents the BH horizon.
     (b) Specific angular momenta of Keplerian orbits for various BS solutions ($k=1$). The solid black line shows $l_{\rm ICO}$ vs $r_{\rm ICO}$ for all of the BS. Thus the colored lines above and below correspond to the prograde and retrograde orbits, respectively. The black region of the colored lines represent unbound orbits, and the magenta regions superluminal orbits.}
     \label{AllLK}
\end{figure}

In order to address the torus topology, a detailed analysis of the Keplerian orbits is necessary. In fact, given a torus endowed with constant specific angular momentum $l_0$, the locations for which $l_{\rm K}^{\pm}(r)=l_0$ can be realized as centers or cusps of the torus. Namely, if the Keplerian specific angular momentum profile is monotonic, only single-centered and cusp-less tori are feasible. The existence of a local minimum \footnote{For the sake of brevity we shall consider the modulus of the Keplerian specific angular momentum profiles, when addressing local maxima and minima.} makes it possible for a cusp to be realized. Finally if a local maximum and a local minimum are both present, double-centered solutions might be possible. In light of this we proceed with a description of the Keplerian orbits around the BSs studied in this work, for which the main result can be found in Fig.~\ref{fig:allK_BS}, a plot of both retrograde and prograde Keplerian orbits specific angular momenta.

\begin{figure}[b]
 \hspace{-1.0cm}\includegraphics[scale=0.50]{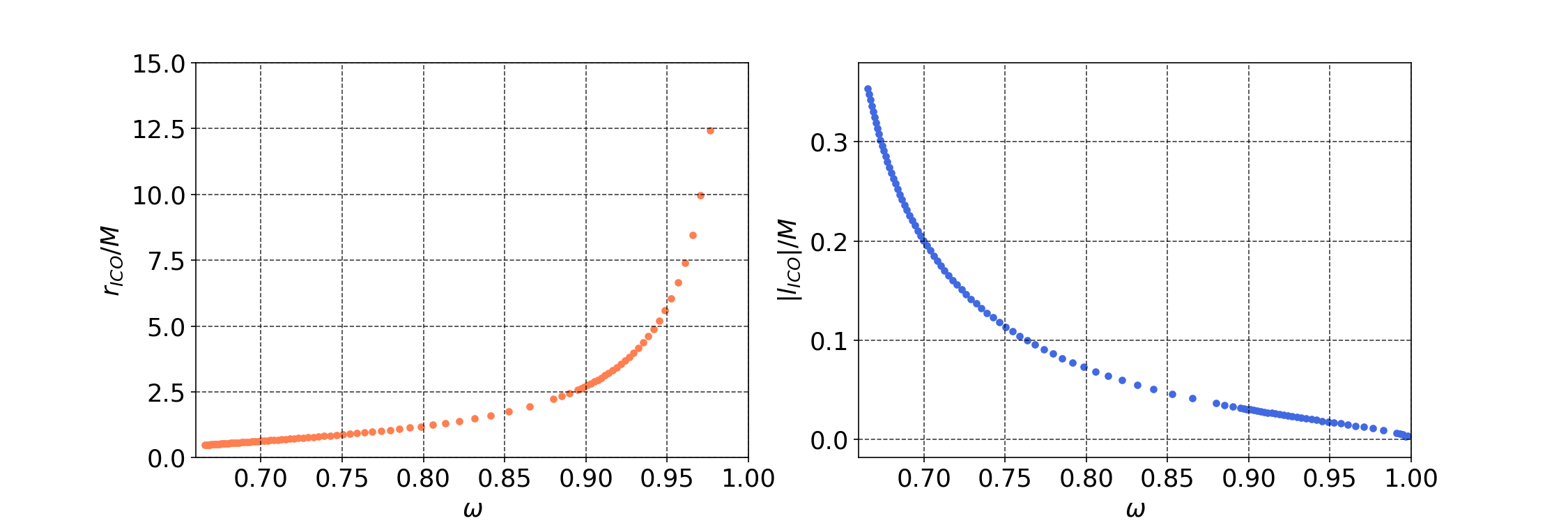}%
     \caption{ICO radii (a) and specific angular momenta (b) for the considered set of BSs ($k=1$).}
     \label{ICO_plot}
\end{figure}

Keplerian orbits exit only for $r>r_{\rm ICO}$. In Fig.~\ref{ICO_plot} the dependence of $r_{\rm ICO}$ and $l_{\rm ICO}$ on $\omega$ is shown. The more relativistic the BSs are, the smaller the ICO radius becomes and the larger the $|l_{\rm ICO}|$. From eq.~(\ref{lpm-Kepler}) it is clear that $l_{\rm K}^+(r_{\rm ICO})=l_{\rm K}^-(r_{\rm ICO}):=l_{\rm ICO}$. For all of the BSs presented here $l_{\rm ICO}<0$. We note that, as observed in \cite{Meliani:2015zta}, $l^+_{K}$ can have negative values close to the ICO.
A plot of both $l_{\rm K}^-$ and $l_{\rm K}^+$ is shown in Fig.~\ref{fig:allK_BS}. For comparison, the prograde and retrograde Keplerian specific angular momenta of a Kerr black hole are shown in Fig.~\ref{fig:kerr}.

The specific angular momenta of the prograde orbits are all monotonic for the set of BSs analyzed. It can also be  observed that these orbits are all subluminal and bound. The absence of local extrema allows these stars to shelter tori with only one center and no cusp. In contrast, retrograde orbits can exhibit a local maximum and a minimum for the solutions with $\omega<0.853$, making it possible for them to house double-centered tori (see section 5). We note that, to our knowledge, before this work only co-rotating tori with similar topologies were found and only around BSs with higher winding numbers \cite{Meliani:2015zta}. We shall address the values of the local extrema of $|l_{\rm K}^-(r)|$ (with $r>r_{\rm ICO}$), $l_{\rm K}^{\max}$ and $l_{\rm K}^{\min}$, when existent \footnote{In order to abbreviate the notation we shall only use $l_{\rm K}^{\min}$ and $l_{\rm K}^{\max}$ for these local minima and maxima of $|l_{\rm K}^-(r)|$, dropping the $\pm$ signs.}. 

\begin{figure}[h]
 \hspace{-1.5cm}\includegraphics[scale=0.5]{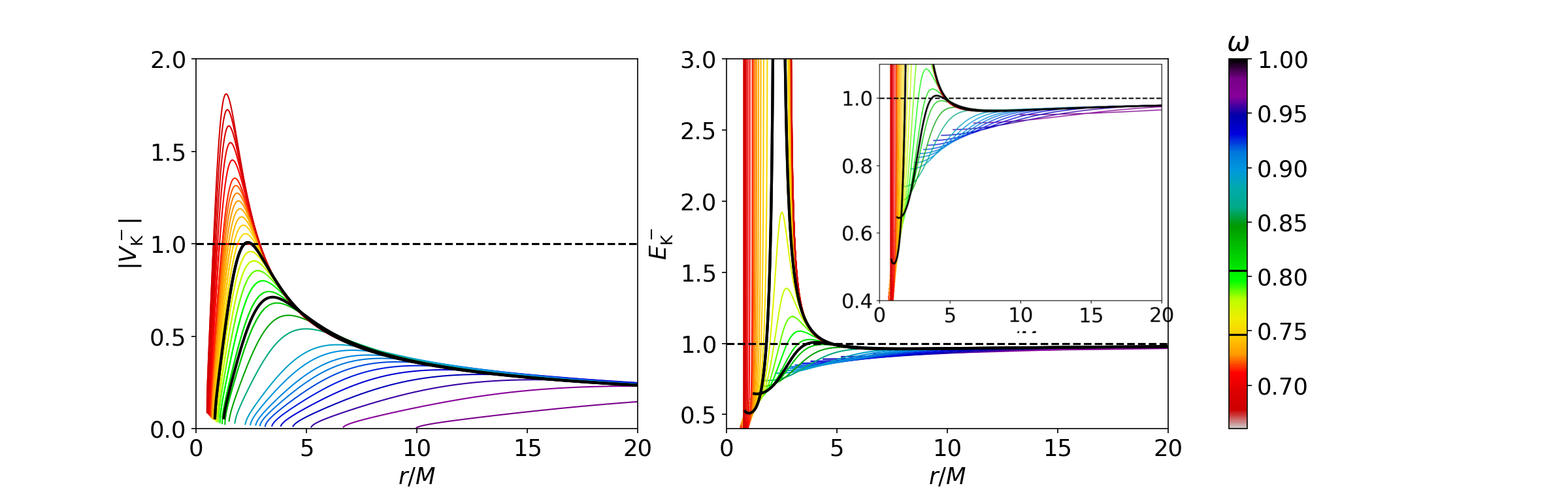}%
     \caption{Specific velocities (a) and energies (b) of Keplerian orbits with respect to a ZAMO. 
     The black curves correspond to the solutions with $\omega=0.806$ and $\omega= 0.746$ and are the thresholds, where regions with $E_{\rm K}^->1$ and $|V_{\rm K}^-|>1$ appear, respectively.}
     \label{AllEV}
\end{figure}

For the retrograde case, unbound orbits ($E_{\rm K}^->1$) are observed for the set of BSs with $\omega\le 0.806$ while superluminal orbits ($|V_{\rm K}^-|>1$) are observed for $\omega \le 0.746$. These orbits' specific angular momenta are depicted in Fig.~\ref{fig:allK_BS} in black and magenta, respectively. Plots of the specific velocities and energies of all orbits can be found in Fig.~\ref{AllEV}. Superluminal regions are found in between the outermost and innermost photon orbits of a given BS ($r_{ph}^{out}$ and $r_{ph}^{in}$), while the unbound orbits appear in between the outermost and innermost marginally bound orbits ($r_{mb}^{out}$ and $r_{mb}^{in}$). In that sense, orbits that are either unbound or superluminal can be found in an annulus with inner radius $r_{mb}^{in}$ and outer radius $r_{mb}^{out}$, noting that $r^{in}_{mb}<r^{in}_{ph}<r^{out}_{ph}<r^{out}_{mb}$. We also note, that $r^{in}_{ph}$ approaches $r_{mb}^{in}$ as $\omega$ decreases, as can be observed in Fig.~\ref{fig:photonorbits}.

In the context of torus building, the role of unbound and superluminal orbits is the following. If a location for which the torus fluid is supposed to be in geodesic motion ($l_{\rm K}^-(r)=l_0$) is inside the annulus of unbound and superluminal orbits, this location will neither be realized as a torus center nor as a cusp in a realistic scenario. This comes from the fact that at these locations $\mathcal{W}$ will be either positive or undefined. In fact, the argument of the logarithm of Eq.~(\ref{potentialW}) will be equal to the specific energy defined in Eq.~(\ref{epm-Kepler}) when the fluid is geodesically moving. The specific energy must then be well defined and smaller than one, in order to accomplish finite tori that include the expected cusp or center. 

It can be seen that the specific angular momentum profiles as well as the existence or nonexistence of marginally bound and photon orbits differ quite dramatically from one BS model to another. In order to address these different scenarios that give rise to distinct torus types, we shall provide a closer look at three of these solutions (section 5). The chosen models are endowed with $\omega=0.960$, $0.768$ and $0.671$. The Keplerian specific angular momentum profiles, the positions of the marginally bound and photon orbits as well as the scalar field profiles of these space-times can be found in Fig.~\ref{Fig:3Solutions}.

\begin{figure}[h]
\centering
 \includegraphics[scale=0.4]{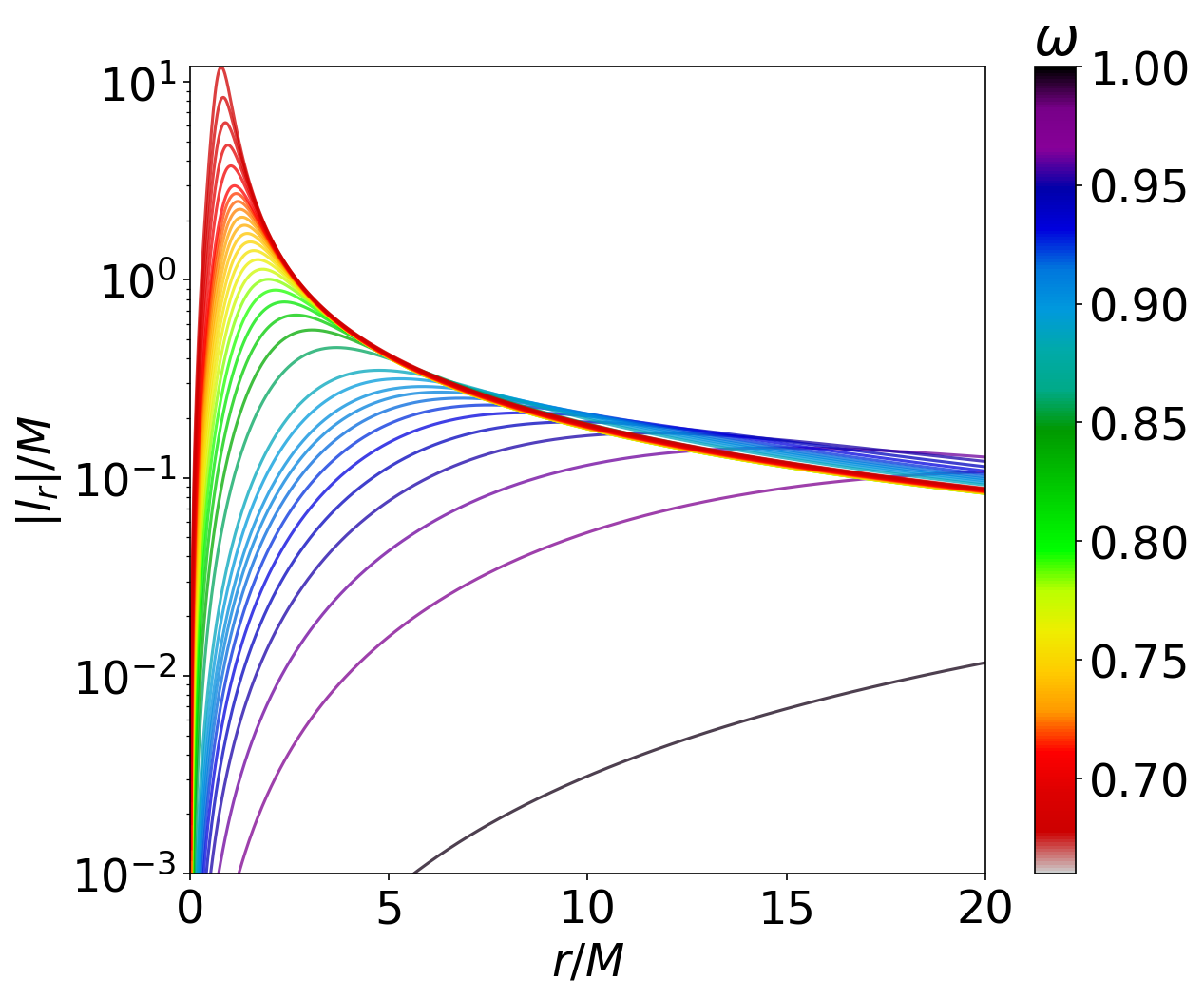}%
     \caption{Rest specific angular momentum profiles for various BS models    with $k=1$ (logarithmic scale).}
     \label{restl:fig}
\end{figure}

\begin{figure}
\centering
\subfloat[ ]{%
  \includegraphics[clip,width=0.5\columnwidth]{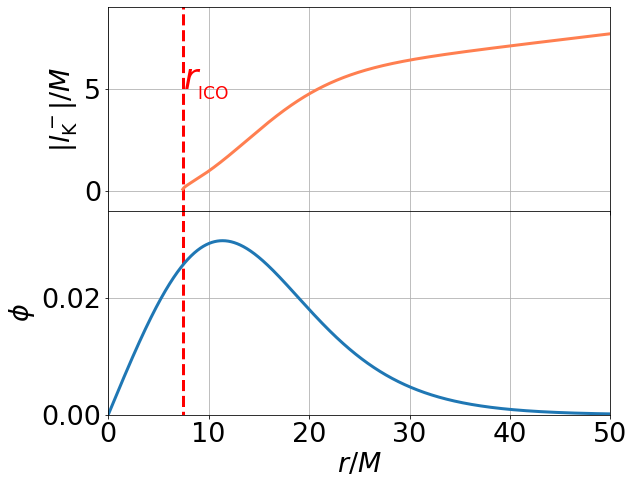}%
 \label{fig:ex1}
}

\subfloat[ ]{%
  \includegraphics[clip,width=0.5\columnwidth]{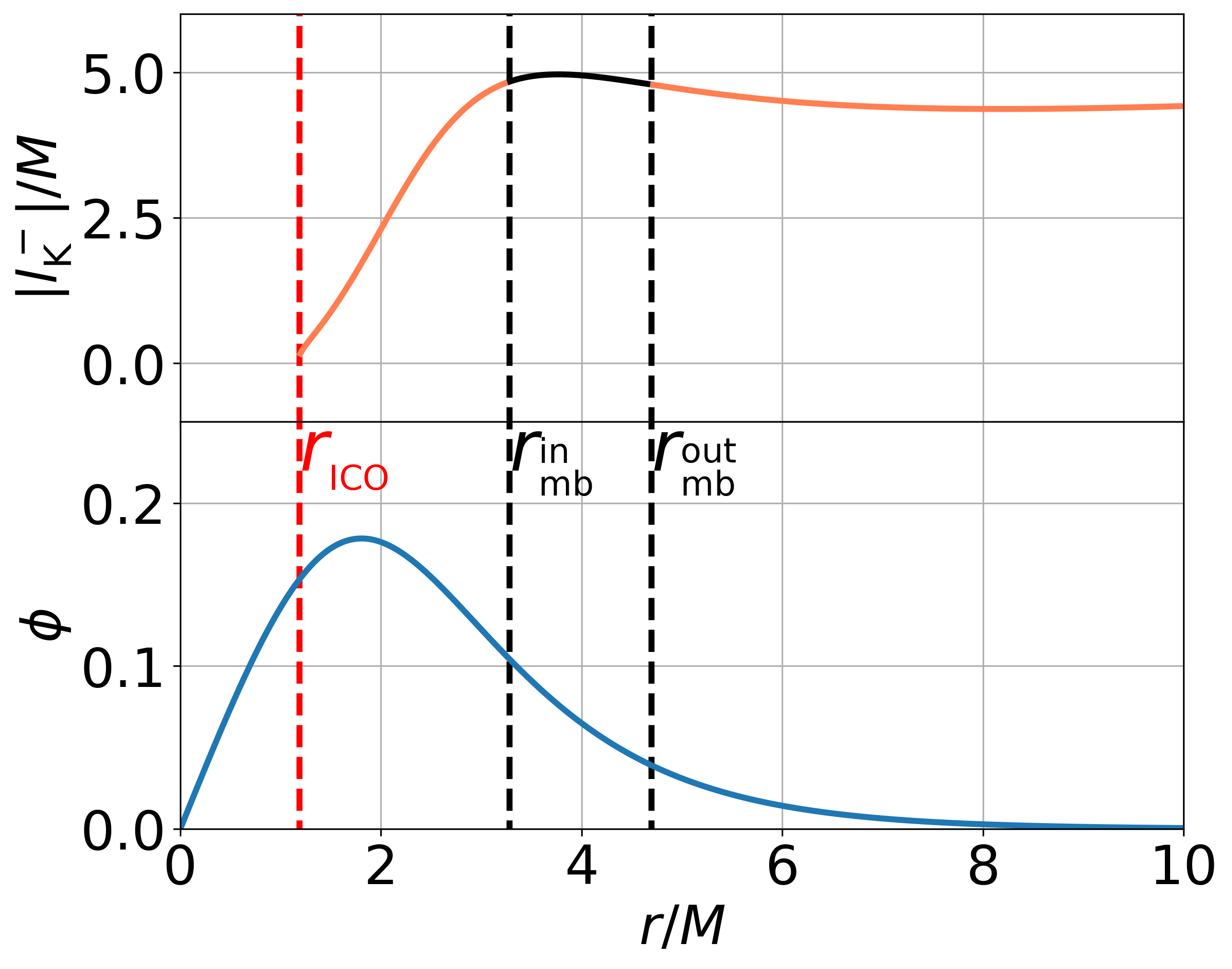}%
\label{fig:ex2}
}

\subfloat[ ]{%
  \includegraphics[clip,width=0.5\columnwidth]{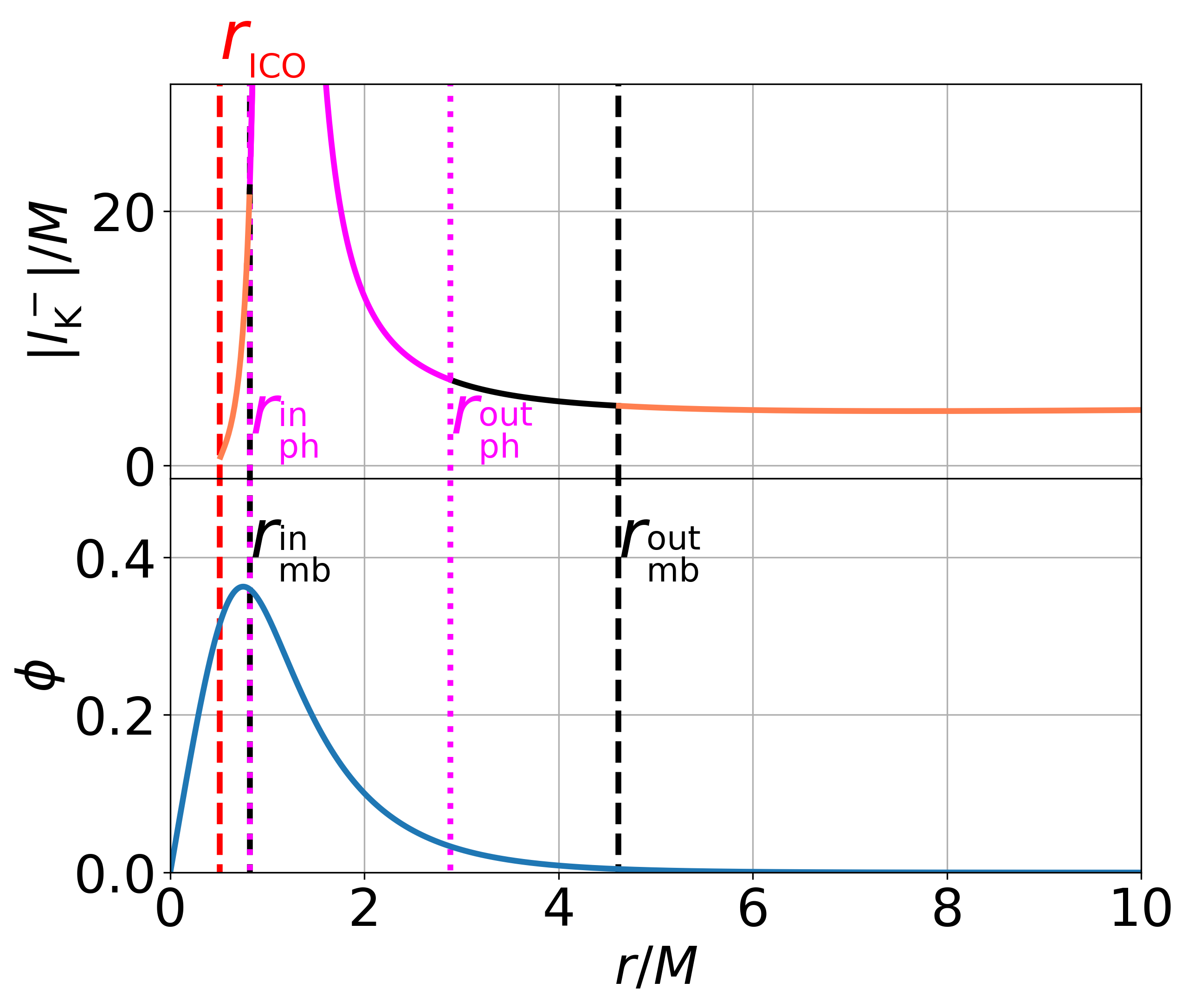}%
  \label{fig:ex3}
}

\caption{(Upper panels) Modulus of the Keplerian angular momentum for three different models of BSs endowed with $\omega=0.960$ (a), $0.798$ (b) and $0.671$ (c) ($k=1$). The position of the ICOs are depicted in red while the positions of the innermost and outermost photon and marginally bound orbits are plotted in magenta and black, respectively. (Lower panels) The scalar fields $\phi$ for the same three solutions.} 
\label{Fig:3Solutions}
\end{figure}

In comparison, for the Kerr metric, both in the retrograde and the prograde case, unbound orbits exist for radii smaller than the location of the marginally bound orbit, and superluminal orbits can be found for radii smaller than the position of the photon orbit  \cite{Bardeen:1972fi}. Thus, no bound subluminal orbits exist between the photon orbit and the event horizon, whereas for BSs these types of orbits can be found between the ICO and the innermost photon orbit. In addition, the Keplerian specific angular momentum profiles are endowed with a local minimum and no local maximum. Thus the tori hosted by the Kerr metric may possess only one center and possibly one cusp (see Fig.~\ref{fig:kerr}).

Another important quantity regarding the torus classification is the rest specific angular momentum defined by Eq.~(\ref{restl}). Given a torus with $l_0$, there might exist a \emph{static surface}, defined by $\mathcal{S}=\{(r,\theta):l_r(r,\theta)=l_0\}$, where the torus angular velocity vanishes, $\Omega(r,\theta)|_{\mathcal{S}}=0$. For the set of BSs we analyzed, $\mathcal{S}$ is always closed. Although the remaining part of the torus is in retrograde motion, inside $\mathcal{S}$ the gas moves in a prograde manner. In order to address the existence of $\mathcal{S}$ we turn our attention to $l_r(r,\pi/2)=l_r(r)$ \footnote{We would like to note that $l_r(r)$ is not necessarily related to a Keplerian orbit. In fact, that is the case only for $l_r(r_s)$, the value at the static ring location.}. In Fig.~\ref{restl:fig} we provide the modulus of this function for our BS models. The condition for a retrograde torus solution to shelter a static surface is then given by  $|l_0|<l_r^{\max}:=\max{(|l_r(r)|)}$. For the more relativistic BSs, $l_r^{\max}$ can assume relatively large values. Thus a large range of $l_0$ will give rise to tori with static surfaces. However, when the value of $\omega$ of the BSs approaches $1$ this range becomes smaller. 
 
Finally we would like to address also the relation between the static orbits and the static surfaces introduced in this work. Static orbits, as described at the end of Sec.~\ref{sec:rbs}, are a special type of circular geodesic in which a particle remains at rest with respect to a ZAMO at infinity. Given the radius of such orbits, $r_{s}$, by definition their specific angular momentum is given by $l_{r}(r_s)$. Since these orbits are also Keplerian, it is possible to find the relation $l_{\rm K}^-(r_s)=l_r(r_s)$, and thus $r_s>r_{\rm ICO}$. It is then always possible, given a BS endowed with a static ring, to choose $l_0$ in such a way that $l_0=l_{\rm K}^-(r_s)=l_r(r_s)$. This means that static surfaces are always feasible for space-times endowed with a static orbit. In fact this particular solution represents the only one,  for a given BS, where one of the torus centers is located at the static ring and, since it moves geodesically at this location, is in a static orbit. Other solutions endowed with static surfaces can have the torus center either inside or outside such surfaces. 

\section{Thick retrograde tori}

\subsection{Classification}

The set of BSs explored in this work can shelter different  types of tori. Therefore we now proceed to classify these fluid solutions by their topology and the presence of static surfaces. We note that all torus solutions described in this and the forthcoming sections address only retrograde tori, thus the condition $l_0<l_{\rm ICO}$ is always assumed. 

From a topological point of view three different types of fluid structures can be formed around these BSs,  classified by their number of feasible centers and cusps. By ``feasible'' we mean that at these torus points the potential $\mathcal{W}$ is negative, and thus these centers/cusps can be realized by finite fluid configurations. We shall enumerate the torus types from $1$ to $3$. Conditions on $l_0$ for these torus types to be realized are shown in Table~\ref{tab:types}.


\begin{table}[h!]
  \begin{center}
 {
    \begin{tabular}{|c|c|c|c|c|c|c|c|c|c|c|c} 
      \hline
     \textbf{\#} &  \textbf{Centers}&  \textbf{Cusp}  & \textbf{$l_0$ condition} & \textbf{BS models} \\
      \hline
      \multirow{3}{*}{Type 1} & \multirow{3}{*}{1} & \multirow{3}{*}{0} & $|l_0|\notin (l_{\rm K}^{min},l_{mb}^{in})$  & $0.665\le\omega\le0.806$\\
            \cline{4-5}
            & & &  $|l_0|\notin (l_{\rm K}^{\min},l_{\rm K}^{\max})$ & $0.806<\omega\le0.853$  \\
            \cline{4-5}
            & & & no condition &  $0.853<\omega<1.000$  \\
      \hline
      \multirow{2}{*}{Type 2 }& \multirow{2}{*}{2} & \multirow{2}{*}{1} & $|l_0|\in \, (l_{\rm K}^{min},l_{mb}^{out})$ &  $0.665\le\omega\le 0.806$\\
      \cline{4-5}
      & & &$|l_0|\in (l_{\rm K}^{min},l_{\rm K}^{\max})$ & $0.806< \omega\le 0.853$  \\
      
      \hline
      Type 3 & 2 & 0 & $|l_0|\,\in\, (l_{mb}^{out},l_{mb}^{in})$  & $0.665<\omega<0.806$\\
      \hline
    \end{tabular}
\caption{Conditions on $l_0$ for the different torus types regarding their topologies (see Fig.~\ref{fig:phase} for a phase diagram). Static surfaces will be present in these tori types if, in addition to the above conditions, $|l_0|<l_{r}^{\max}$.}
\label{tab:types}
   }
\end{center}
\end{table}

\emph{Type 1} tori are the simplest and most abundant, having only one feasible center and no cusp. \emph{Type 2} tori are double-centered and can possess a cusp, with their innermost center residing typically close to the $r_{\rm ICO}$ and possessing a higher density than their outermost center. \emph{Type 3} tori, while also double-centered can not be endowed with a cusp. Thus the centers of this torus type are disconnected. No torus solutions endowed with only one center and one cusp can be found for this set of BSs, in contrast to the Kerr case.

As seen in Table \ref{tab:types},  the conditions on $l_0$ with respect to the possible torus topologies depend on the BS model. Due to the diversity of the solutions, these conditions may differ considering two factors: the presence of marginally bound orbits and the monotonicity of the Keplerian specific angular momentum for a given BS. In this sense we can divide the BS solutions into three sectors. 

\begin{itemize}
\item ($0.665\le\omega\le 0.806$) Non-monotonic $l_{K}^-(r)$ and two marginally bound orbits .
\item ($0.806<\omega\le 0.853$) Non-monotonic $l_{K}^-(r)$ and no marginally bound orbits.
\item ($0.853<\omega<1$.000)  Monotonic $l_{K}^-(r)$ and no marginally bound orbits.
\end{itemize}

Found for BSs with $0.665<\omega<0.806$, marginally bound orbits indicate the presence of a region in which neither cusps nor centers can be realized, regardless if the  condition $l_{\rm K}^-(r)=l_0$ is satisfied.  Such a region shelters superluminal and unbound Keplerian orbits, and it is delimited by $r_{mb}^{in}$ and $r_{mb}^{out}$. Thus, for the BSs endowed with such a region, the relation between $l_0$ and the torus topology must now take into account the specific angular momentum of the innermost and outermost marginally bound orbits, $l_{mb}^{in}:=|l_{\rm K}^-(r_{mb}^{in})|$ and $l_{mb}^{out}:=|l_{\rm K}^-(r_{mb}^{out})|$, respectively \footnote{It can be observed that $l_{mb}^{in}>l_{mb}^{out}>l_{\rm K}^{\min}$.}. Solutions endowed with only a  single center will be realized by choosing $|l_0|$ outside the interval $(l_{\rm K}^{\min},l_{mb}^{in})$.  Meanwhile if $|l_0|\in(l_{\rm K}^{\min},l_{mb}^{out})$, the condition $l_{\rm K}^-(r)=l_0$ will be satisfied at three points outside the unbound/superluminal regions, yielding a torus endowed with two centers and a cusp. On the other hand, if $|l_0|\in(l_{mb}^{out},l_{mb}^{in})$, the condition $l_{\rm K}^-(r)=l_0$ will be satisfied also for three radii. Two of these locations shall represent feasible centers, whereas the position usually related to the cusp will now correspond to an unbound or superluminal orbit. In the former case $\mathcal{W}(r_{cusp})>0$, while in the later $\mathcal{W}(r_{cusp})\rightarrow\infty$. Thus this double-centered configuration will be cusp-less.
\begin{figure}[h]
\centering
 \includegraphics[scale=0.50]{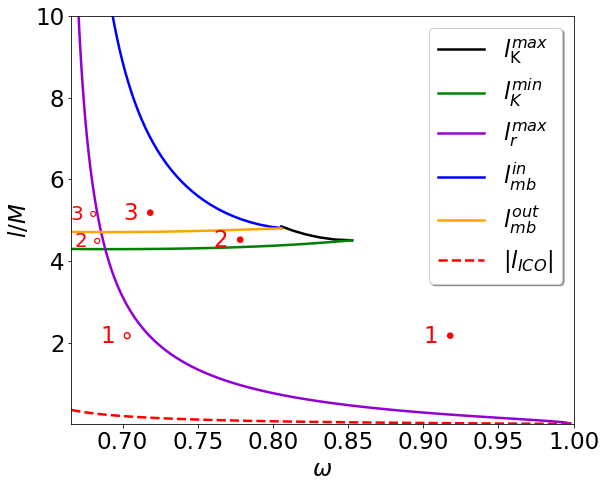}%
     \caption{Phase diagram of $|l_0|$ vs $\omega$, with all types of torus solutions. $l_{\rm K}^{\max}$ is plotted only for BSs not endowed with marginally bound orbits. The red numbers correspond to the torus types described in Table~\ref{tab:types}, while the symbols $\bullet$ and $\circ$ correspond to the absence and presence of a static surface, respectively.}
     \label{fig:phase}
\end{figure}

 For $0.806<\omega\le 0.853$, the Keplerian  specific angular momentum features a local maximum and minimum, but neither superluminal nor unbound orbits. Only Type 1 and Type 2  tori are feasible in these space-times and can be realized by taking $|l_0|\notin (l_{\rm K}^{\min}, l_{\rm K}^{\max})$ or $|l_0|\in(l_{\rm K}^{\min}, l_{\rm K}^{\max})$, respectively. For the BSs with boson frequency larger than $0.853$, the Keplerian specific angular momentum increases monotonically with the radial coordinate $r$. Thus only Type 1 tori can be sheltered in these space-times, regardless of the choice of $l_0$.  

 Furthermore these three torus types can also be split into two categories regarding their static surfaces. Hereafter we shall add the symbols $\circ$ and $\bullet$ behind the type numbers to indicate the presence or absence of static orbits for a given solution, respectively.  Tori for which $|l_0|<l_r^{\max}$ will be endowed with a static surface, whereas all others will not. For all the BSs presented in this work, tori endowed with a static surface are possible.
As it can be seen in Fig.~\ref{restl:fig}, the value of $l_r^{\max}$ increases as the boson frequency of the solutions, $\omega$, decreases. Hence, the most relativistic BSs are inclined to shelter static surfaces for a wider range of $l_0$, when compared with the less relativistic ones.

A complete phase diagram of the feasible tori can be found in Fig.~\ref{fig:phase}. 

To illustrate these torus types and clarify how these conditions on $l_0$ work, we shall proceed by providing specific solutions of tori around three distinct BSs from the analysed set. 

\subsection{Examples of torus solutions}

\subsubsection{$\omega=0.960$}

We consider first a BS with $\omega=0.960$ and mass $M=0.780$. The ICO is at $r_{\rm ICO}=7.406M$ with specific angular momentum $l_{\rm ICO}=-0.0147M$. The Keplerian specific angular momentum modulus $|l_{\rm K}^-(r)|$ increases monotonically with $r$. Therefore this BS can only shelter tori endowed with a single center  (Type 1). 
The function $|l_{\rm K}^-(r)|$ as well as the scalar field $\phi(r)$ are depicted in Fig.~\ref{fig:ex1}. Neither unbound nor superluminal orbits are present in this space-time.

\begin{figure}[h]
\centering
\begin{subfigure}{.5\textwidth}
  \centering
  \includegraphics[width=1\linewidth]{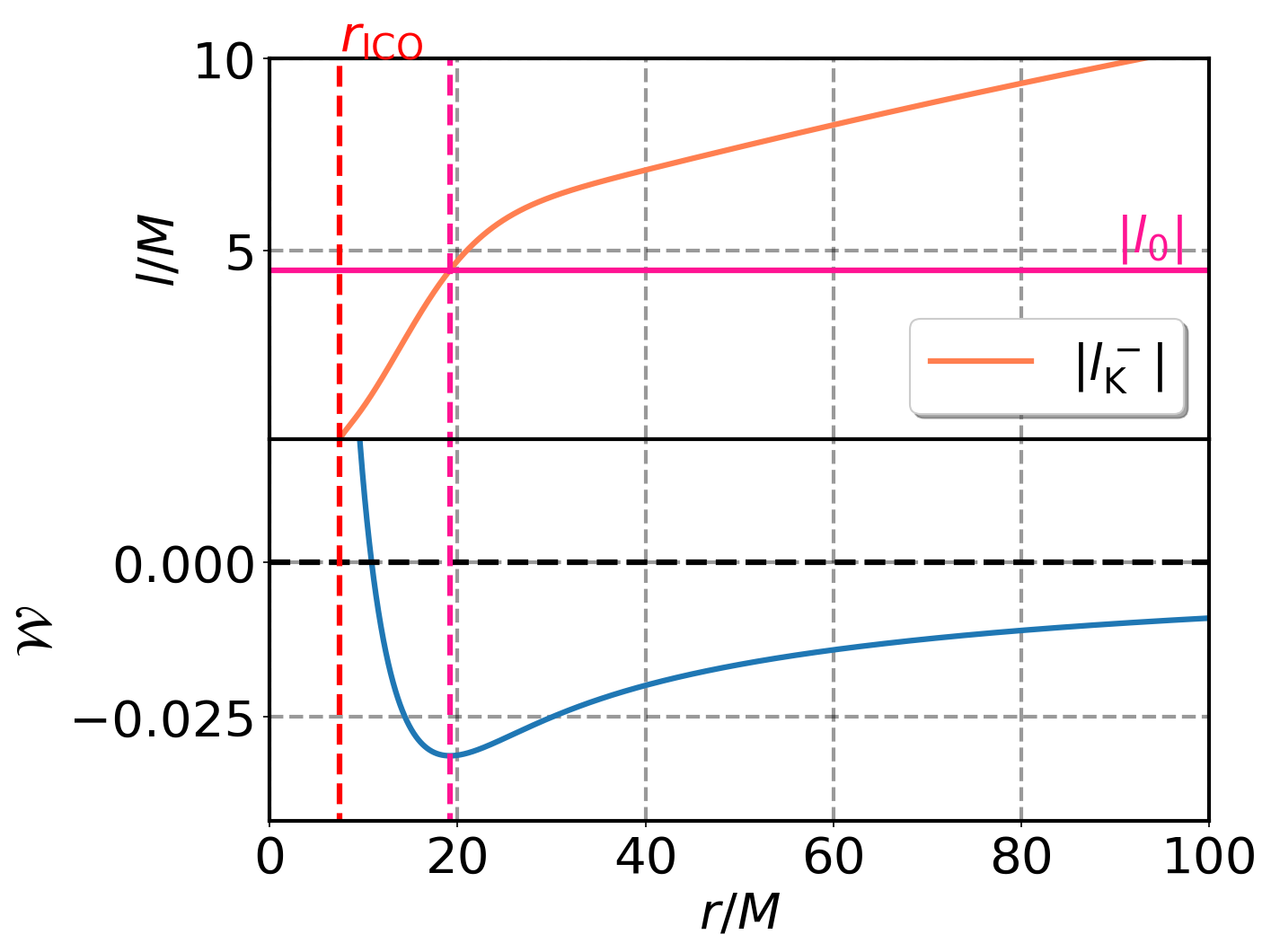}
  \caption{ }
  \label{fig:pot_om0960_l45}
\end{subfigure}%
\begin{subfigure}{.5\textwidth}
  \centering
  \includegraphics[width=1\linewidth]{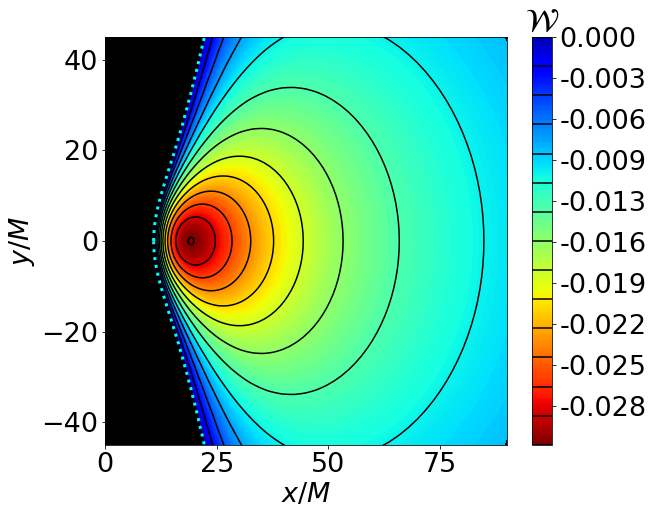}
 \caption{ }
  \label{fig:eqpot_om0960_l45}
\end{subfigure}
\caption{Example of Type 1$\bullet$ torus: (a - upper panel) Retrograde Keplerian specific angular momentum  for a BS with $\omega=0.960$ and $k=1$. (a - lower panel) Potential for  a torus with $l_0=-4.5M$  in the equatorial plane. The dashed magenta line marks the torus center. (b) Equipotential surfaces for the same potential (meridional cross section). The dotted bright blue line represents the surface $\mathcal{W}(r,\theta)=0$.}
\label{fig:om0960_l45}
\end{figure}

An example of a Type 1$\bullet$ torus can be built by choosing $l_0=-4.5M$. The potential $\mathcal{W}(r)$ and the equipotential surfaces for $\mathcal{W}(r,\theta)$ are shown in Fig.~\ref{fig:om0960_l45}. The torus is centered at $r_{\rm center}=19.2M$.  

\begin{figure}[h]
\centering
\begin{subfigure}{.5\textwidth}
  \centering
  \includegraphics[width=1\linewidth]{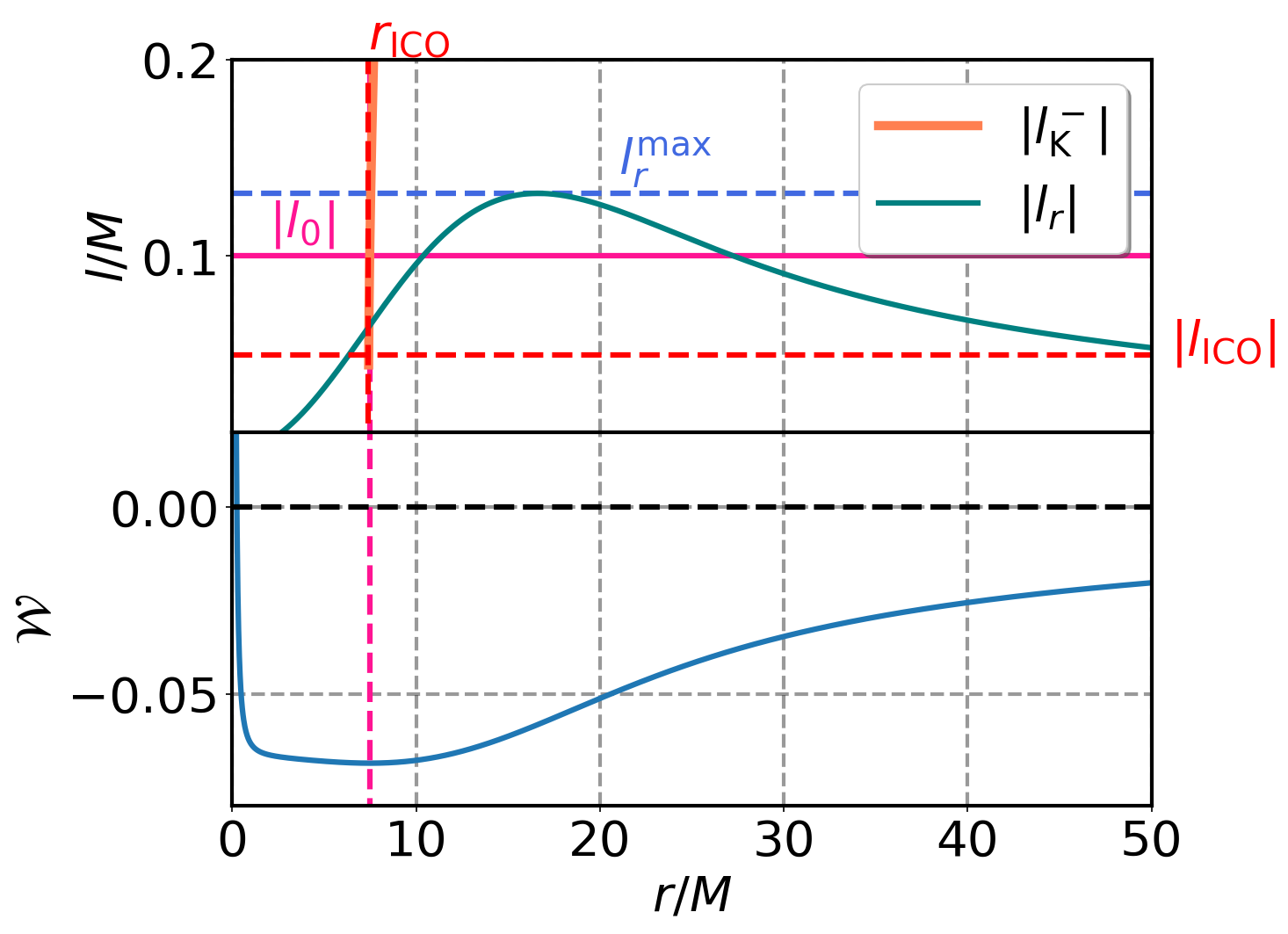}
  \caption{ }
  \label{fig:pot_om0960_l01}
\end{subfigure}%
\begin{subfigure}{.5\textwidth}
  \centering
  \includegraphics[width=1\linewidth]{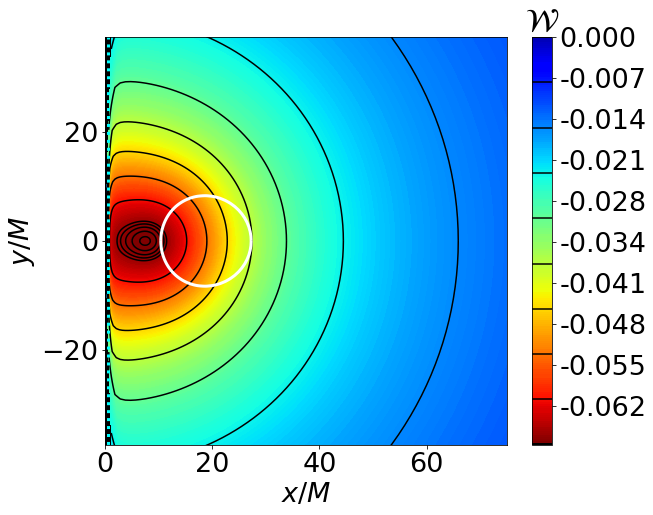}
 \caption{ }
  \label{fig:eqpot_om0960_l01}
\end{subfigure}
\caption{Example of Type 1$\circ$ torus: (a - upper panel) Retrograde Keplerian specific angular momentum  for a BS with $\omega=0.960$ and $k=1$. (a - lower panel) Potential for a torus with $l_0=-0.1M$ in the equatorial plane. The dashed magenta line marks the torus center (in this case the center is very close to $r_{\rm ICO}$) . (b) Equipotential surfaces for the same potential (meridional cross section). The dotted bright blue line represents the surface $\mathcal{W}(r,\theta)=0$ and the solid white line the static surface.}
\label{fig:om0960_l01}
\end{figure}

The rest specific angular momenta have a maximum $l_r^{\max}=0.132$, thus only a restricted range of $l_0$ is capable of generating tori with static surfaces ($0.0147M<|l_0|<0.132M$). As seen in Fig.~\ref{fig:pot_om0960_l01} these solutions (between the dashed blue and red lines) are fated to be centered extremely close to the ICO, since $r_{\rm center}$ must be at the intersection of $l_0$ and $l_{\rm K}^-(r)$. The static surfaces would typically not contain the torus center.

By lowering the values of $|l_0|$, the torus solutions will be centered closer to the ICO and start to present a static surface. As an example we provide a solution with $l_0=-0.1M$. This torus is centered at $r_{\rm center}=7.48M$. As seen in Fig.~\ref{fig:om0960_l01}, the potential in the equatorial plane increases rather abruptly near the BS center. Although the torus center is relatively far from the BS center, the equipotential surfaces come quickly closer to the BS center, leading to deviations from the typical torus shape, and making the torus extremely thick. In fact, this torus represents an example of a so-called fat torus, where the fluid would be able to fill almost all of the region around the BS. A static surface is located in a region outside $r_{\rm center}$ with a toroidal shape (Fig.~\ref{fig:eqpot_om0960_l01}).

\subsubsection{$\omega=0.798$}

We turn now our attention to an intermediate BS, with $\omega=0.798$ and $M=1.316$. The specific angular momenta of the retrograde Keplerian orbits are shown in Fig.~\ref{fig:ex2}, together with the marginally bound orbit positions and the scalar field $\phi(r)$. The region in between the marginally bound orbits is endowed with unbound orbits (plotted in black), but superluminal orbits do not occur. A local maximum and a local minimum are also present in $l_{\rm K}^-(r)$.  

\begin{figure}
\centering

\begin{subfigure}{.7\textwidth}
  \centering
  \includegraphics[width=1\linewidth]{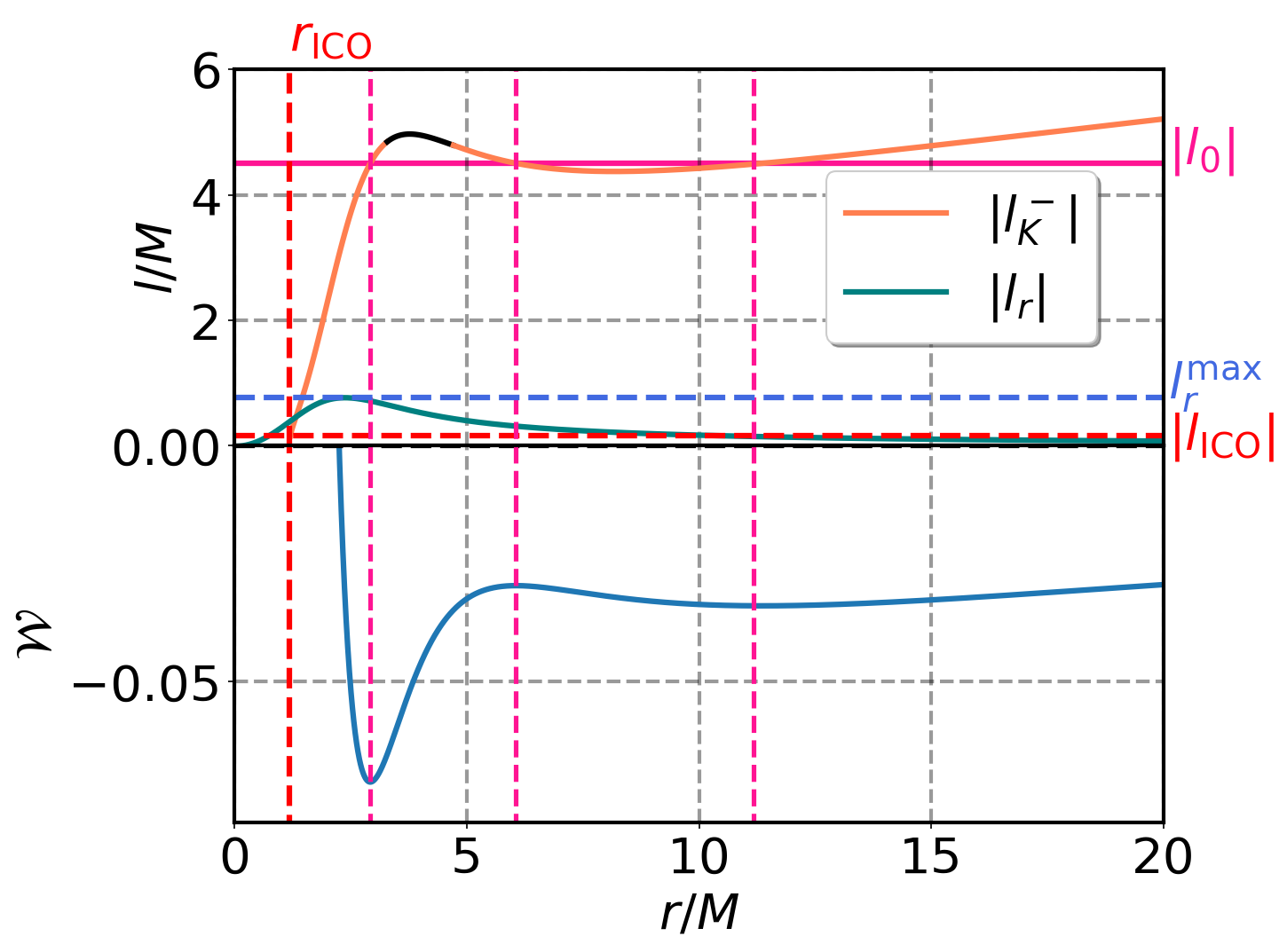}
  \caption{ }
   \label{fig:pot_om_0798_lk45}

\end{subfigure}

\begin{subfigure}{.55\textwidth}
  \centering
  \includegraphics[width=1\linewidth]{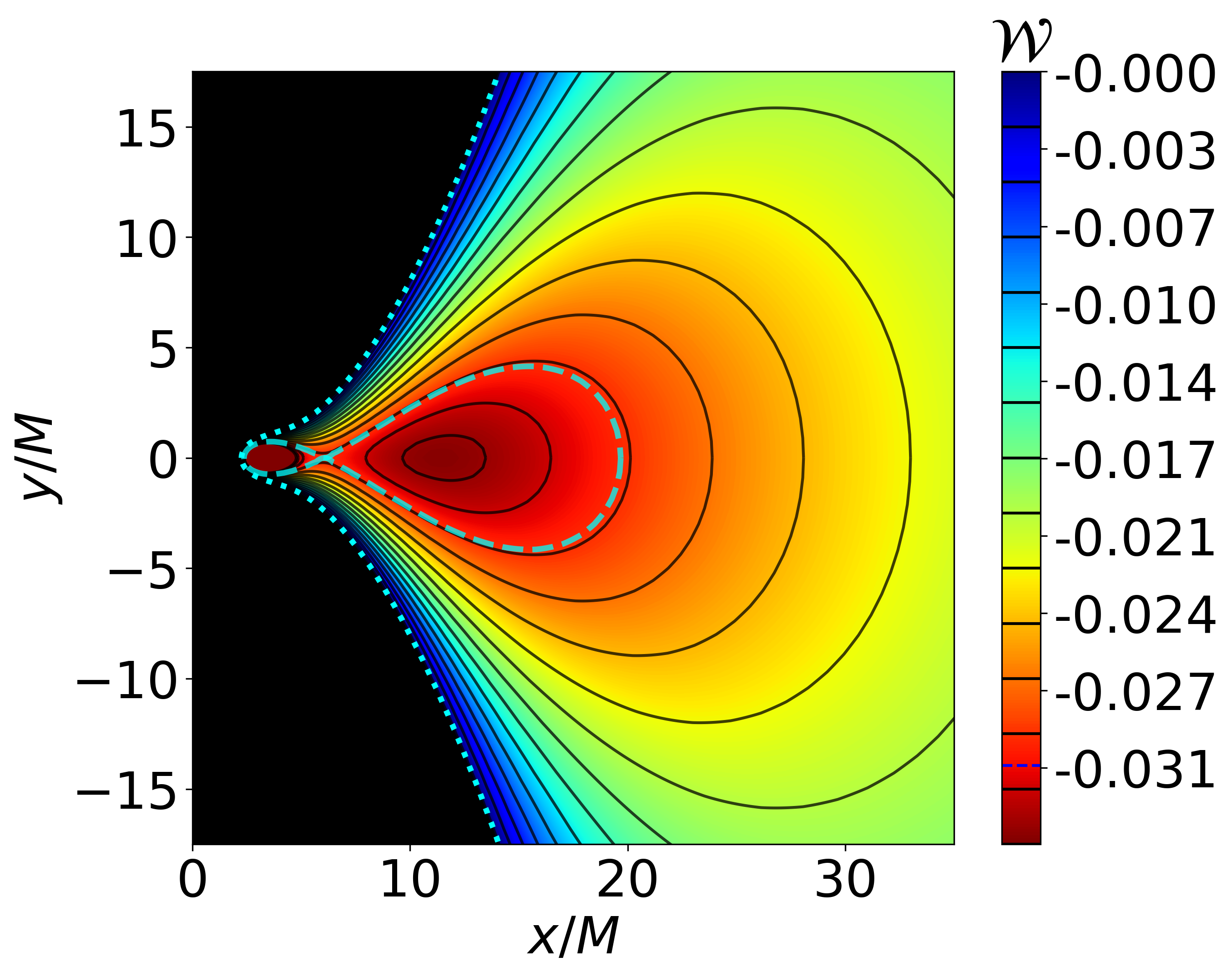}
  \caption{ }
  \label{fig:epot_om_0798_lk45}

\end{subfigure}%
\begin{subfigure}{.55\textwidth}
  \centering
  \includegraphics[width=1\linewidth]{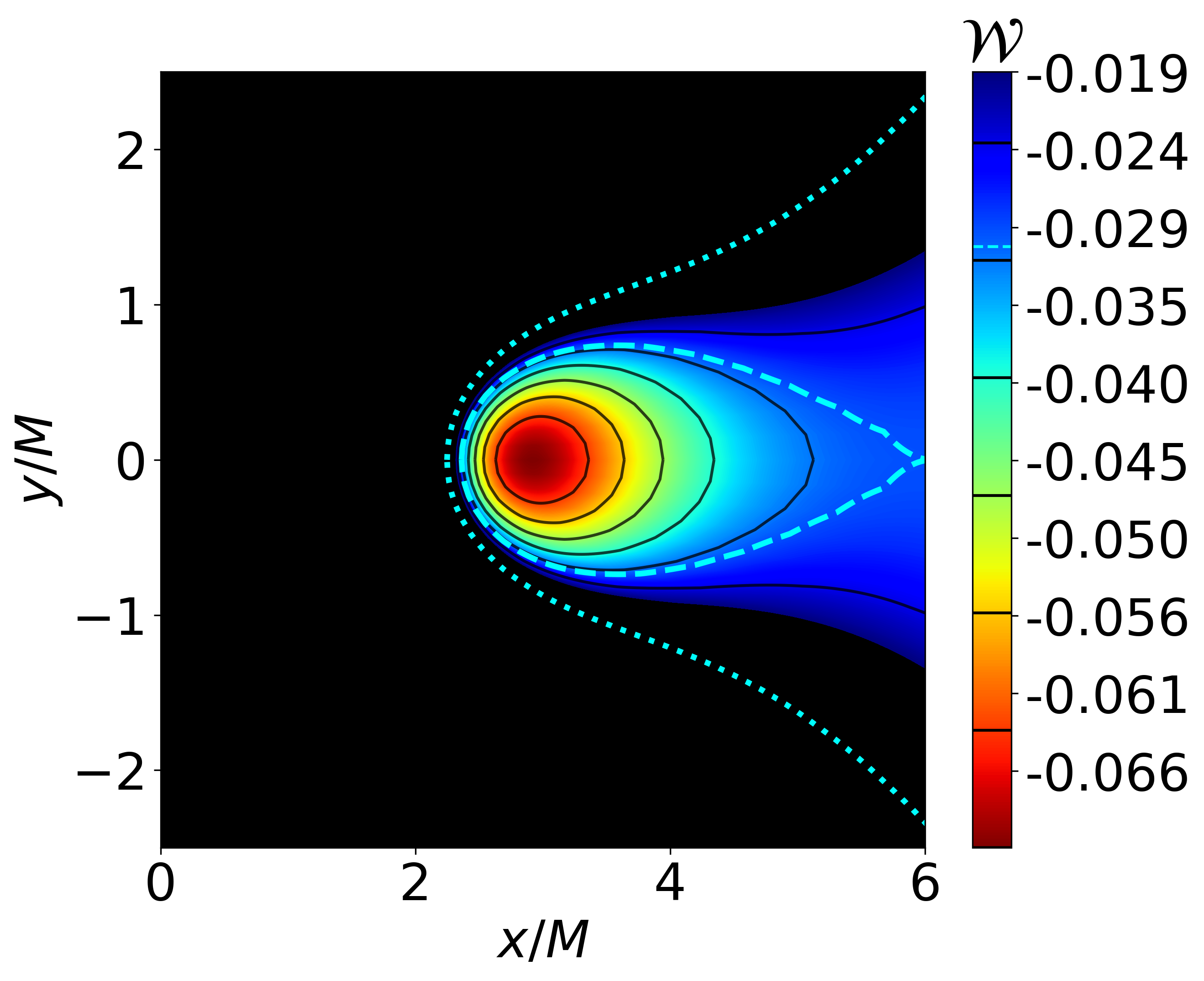}
 \caption{ }
   \label{fig:epot_om_0798_lk45_2}

\end{subfigure}\\

\caption{Example of a Type 2$\bullet$ torus: (a - upper panel) Retrograde Keplerian specific angular momentum  for a BS with $\omega=0.798$ and $k=1$. (a - lower panel) Potential for a torus with $l_0=-4.5M$ in the equatorial plane. The dashed magenta line marks the torus centers and cusp. (b) Equipotential surfaces for the same potential (meridional cross section). The dotted bright blue line represents the surface $\mathcal{W}(r,\theta)=0$, while the dashed bright blue line marks the self-intersecting equipotential related to the cusp.  (c) Same as (b) but for a different range of $\mathcal{W}$ (see color maps) and with a different scale.}

\label{fig:eqp_om0798_l45}
\end{figure}

An example of a Type 2$\bullet$ torus can be generated by setting $l_0=-4.5M$. As expected, the potential is endowed with two local minima corresponding to the two torus centers. The centers of this torus are located at $r_{\rm center}^{(1)}=2.92M$ and $r_{\rm center}^{(2)}=11.37M$. The cusp position is at $r_{\rm cusp}=6.06M$, where a potential local maximum can be found. A depiction of the characteristics of this torus solution can be found in Fig.~\ref{fig:eqp_om0798_l45}.

The equipotential surfaces are shown in Figs.~\ref{fig:epot_om_0798_lk45} and \ref{fig:epot_om_0798_lk45_2}. The self-intersecting equipotential related to the cusp is shown by a dashed bright blue line. It represents a connection between the two regions of the torus. Indeed, when the outer layer of the torus, defined by $\mathcal{W}_{in}$, reaches $\mathcal{W}_{\rm in}=\mathcal{W}(r_{\rm cusp})$ the two sectors of the torus become connected, whereas when $\mathcal{W}_{\rm in}<\mathcal{W}(r_{\rm cusp})$ the two sectors become completely disconnected. 

By relating $\mathcal{W}^{(1)}$ and $\mathcal{W}^{(2)}$ it is possible to infer, for a polytropic fluid, the ratio between the densities of the two centers
\begin{align}
\frac{\rho^{(1)}}{\rho^{(2)}}={\Big(\frac{\exp({-\mathcal{W}^{(1)}})-1}{\exp({-\mathcal{W}^{(2)}})-1}\Big)}^{1/(\Gamma-1)} \, ,
\end{align}
where superscripts $(1)$ and $(2)$ represent the quantities evaluated at the inner and outer torus centers, respectively. For a polytropic index of $\Gamma=4/3$, we observe that $\frac{\rho^{(1)}}{\rho^{(2)}}=9.8$. Thus the density is higher at the innermost torus center than at the outermost torus center. In fact, we find that a higher density at the innermost center is a common characteristic of double-centered tori. 

\begin{figure}
\centering
\begin{subfigure}{.5\textwidth}
  \centering
  \vspace{0.5cm}
  \includegraphics[width=1\linewidth]{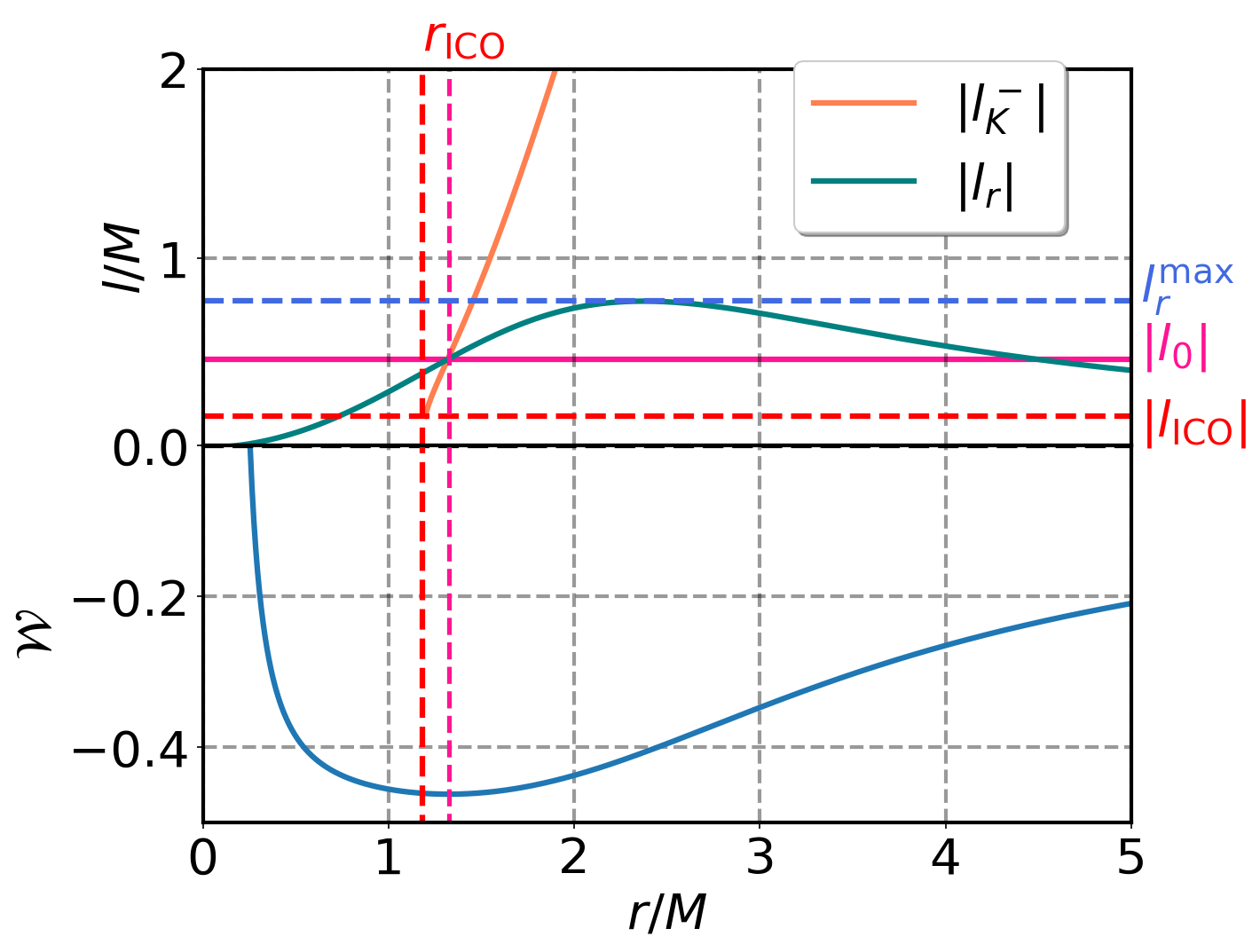}
  \caption{ }
  \label{fig:pot_om_0798_lk0456}
\end{subfigure}%
\begin{subfigure}{.5\textwidth}
  \centering
  \includegraphics[width=1\linewidth]{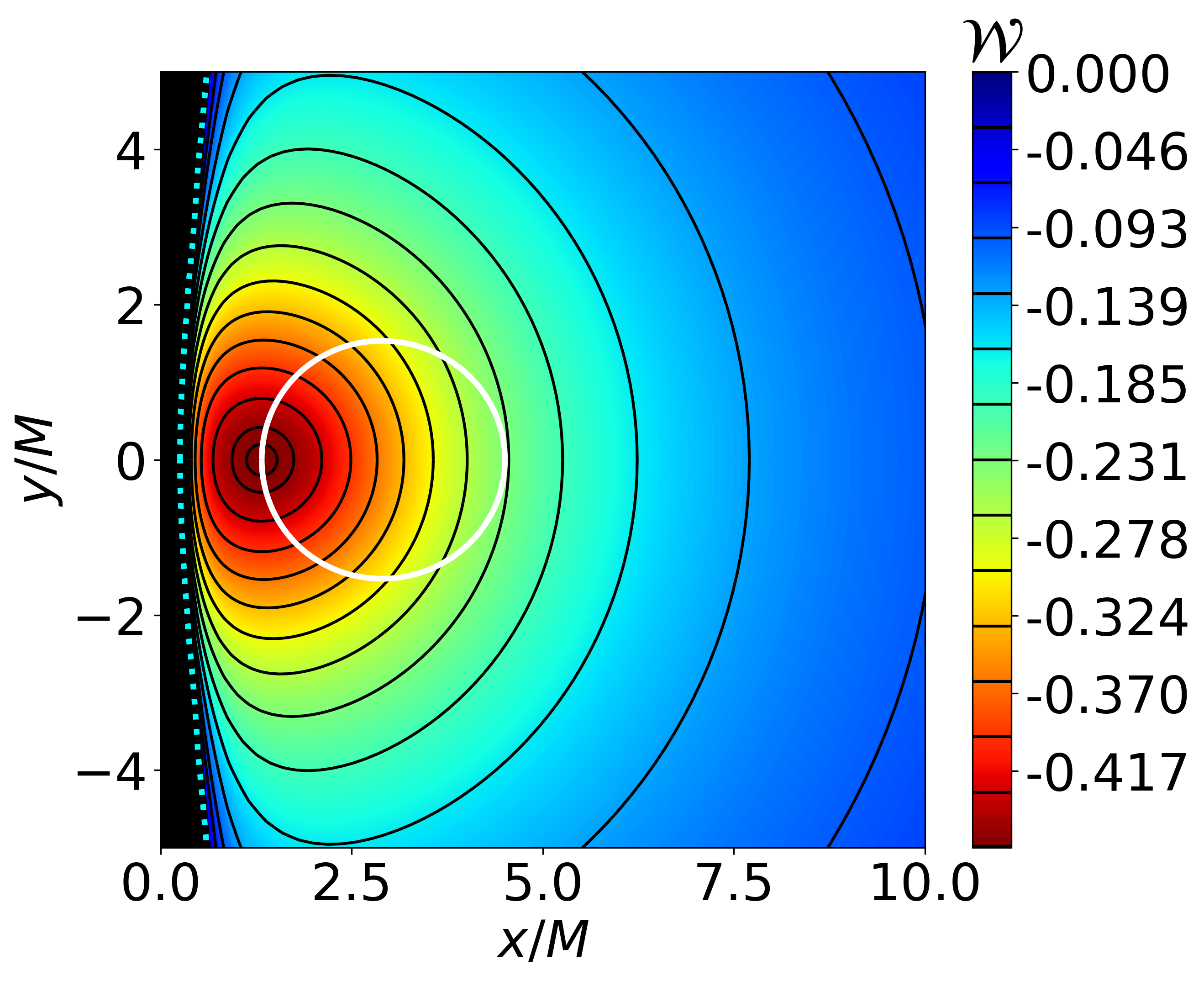}
 \caption{}
 \label{fig:ept_om0798_l046}
\end{subfigure}
\caption{Example of Type 1$\circ$  torus: (a - upper panel) Retrograde Keplerian specific angular momentum for a BS with $\omega=0.798$ and $k=1$. (a - lower panel) Potential for a torus with $l_0=-0.4666M$ in the equatorial plane. The dashed magenta line marks the torus center (in this case the center coincides with the static ring location). (b) Equipotential surfaces for the same potential (meridional cross section). The dotted bright blue line represents the surface $\mathcal{W}(r,\theta)=0$ and the solid white line the static surface.}
     \label{fig:potom0798l_0468}
\end{figure}

Type 1$\circ$ tori can be realized by setting $|l_0|<l_{r}^{\max}=0.776M$. As anticipated in Section 4, static surfaces can intercept the center of the torus since it is possible to choose $l_0=l_{\rm K}^-(r_s)=l_{r}(r_s)$, or in other words, when the torus is centered at the static ring.
This type of solution is also possible for $\omega=0.960$, but we note that as $\omega$ comes closer to $1$ the static ring comes closer to the ICO, as seen in Fig.~\ref{fig:srbs}. 
For the torus center to be inside the static surface it is required that $r_{\rm ICO}<r_{\rm center}<r_{s}$. Therefore it is possible to infer that tori centered inside the static surface are more likely to be sheltered by the more relativistic BSs. In fact, as seen in Fig.~\ref{fig:pot_om_0798_lk0456} (upper panel), for $\omega=0.798$, tori solutions endowed with a center inside the static surface are more abundant as compared to the previous BS ($\omega=0.960$).

As an example we build a torus with $l_0=-0.4666M$, the specific angular momentum required in order to have the torus centered at a point of the static surface. The potential in the equatorial plane as well as the equipotential surfaces and the location of the static surface are shown in Fig.~\ref{fig:ept_om0798_l046}. Interestingly this solution represents yet again another fat torus. 

\subsubsection{$\omega=0.671$}

\begin{figure}

\begin{subfigure}{.5\textwidth}
  \centering
  \includegraphics[width=1\linewidth]{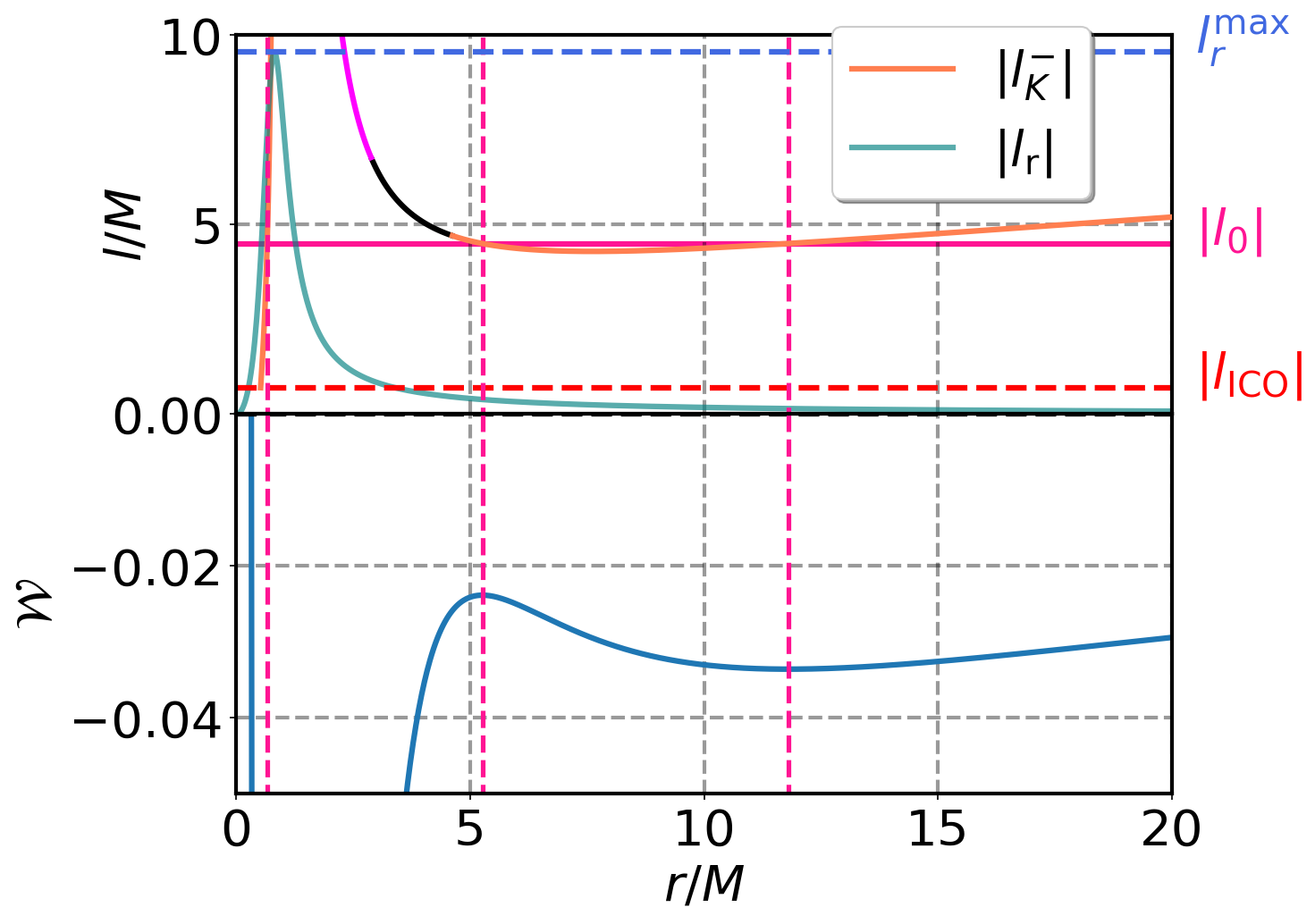}
  \caption{ }
   \label{fig:pot_om_0671_lk45}
\end{subfigure}
\begin{subfigure}{.5\textwidth}
  \centering
  \includegraphics[width=1\linewidth]{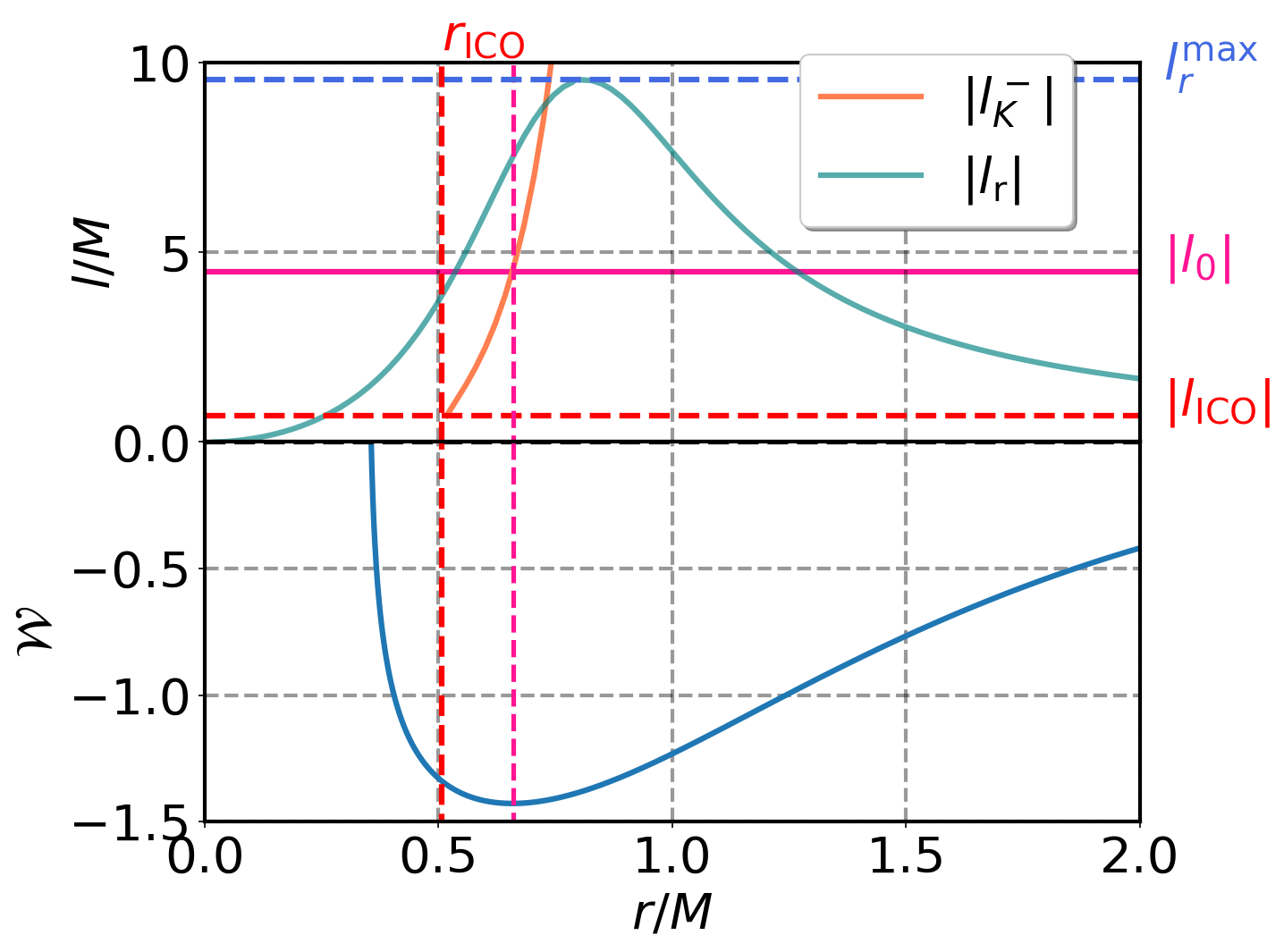}
  \caption{ }
   \label{fig:pot_om_0671_lk45_2}
\end{subfigure}

\begin{subfigure}{.55\textwidth}
  \centering
  \includegraphics[width=1\linewidth]{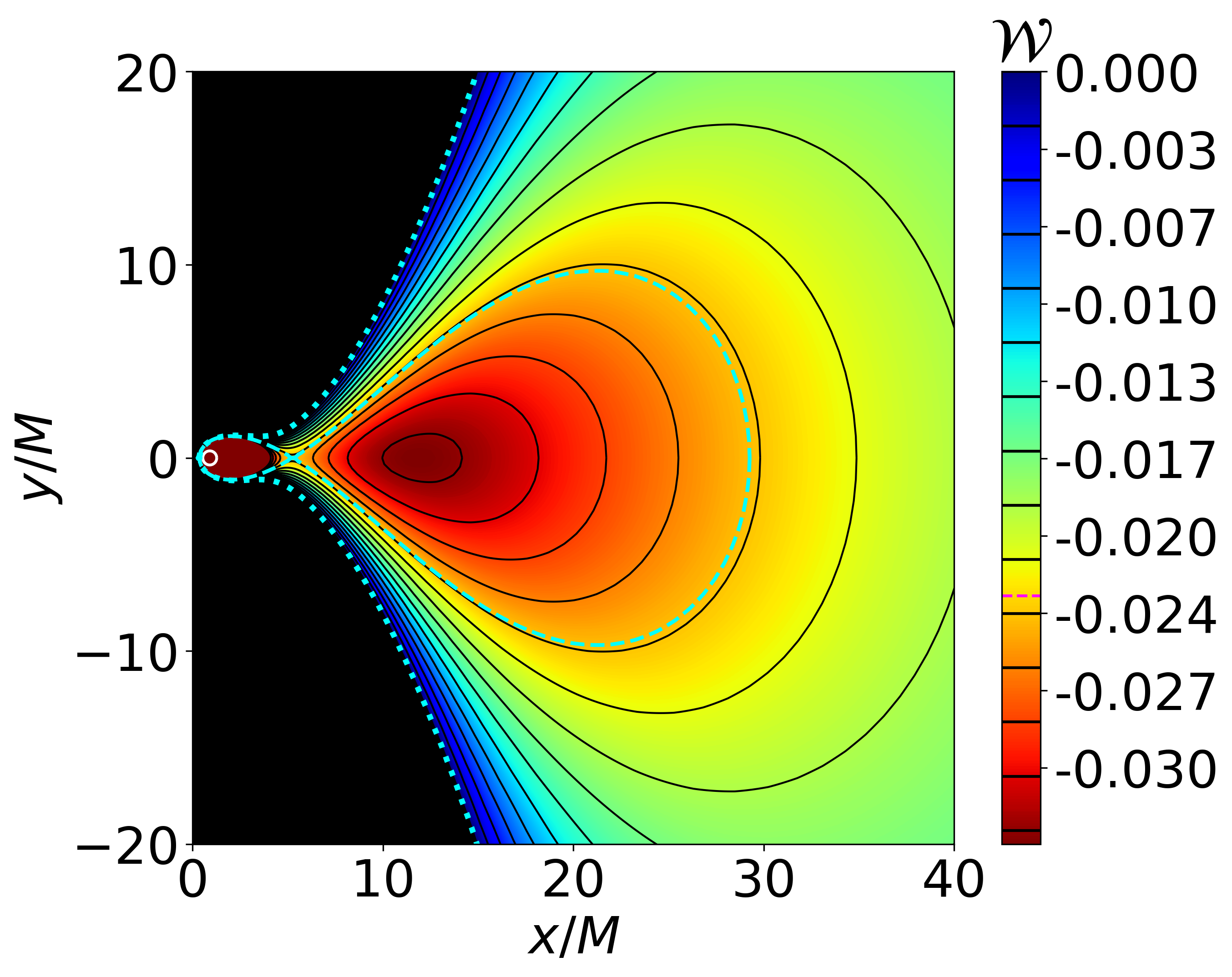}
  \caption{ }
  \label{fig:epot_om_0671_lk45}

\end{subfigure}%
\begin{subfigure}{.55\textwidth}
  \centering
  \includegraphics[width=1\linewidth]{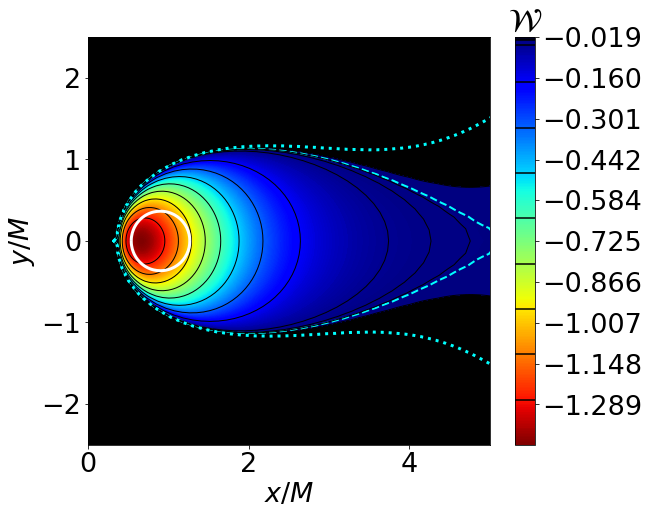}
 \caption{ }
   \label{fig:epot_om_0671_lk45_2}

\end{subfigure}\\

\caption{Example of a Type 2$\circ$ torus: (a - upper panel) Retrograde Keplerian specific angular momentum for a BS with $\omega=0.671$ and $k=1$. (a - lower panel) Potential for a torus with $l_0=-4.5M$ in the equatorial plane. The dashed magenta line marks the torus centers and cusp. (b) Same as (a) but with a different scale, highlighting the center of the innermost torus. (c) Equipotential surfaces for the same potential (meridional cross section). The dotted bright blue line represents the surface $\mathcal{W}(r,\theta)=0$, while the dashed bright blue line marks the self-intersecting equipotential related to the cusp.  (d) Same as (c) but for a different range of $\mathcal{W}$ (see color maps) and with a different scale.}

\label{fig:eqp_om0671_l45}
\end{figure}

Considering now the BS solution with $\omega=0.671$ and $M=1.207$, we observe a much bigger range of retrograde torus solutions endowed with a static surface, since $l_r^{\max}=9.537M$. The Keplerian specific angular momentum distribution can be seen in Figs.~\ref{fig:pot_om_0671_lk45} and \ref{fig:pot_om_0671_lk45_2}. In contrast to the case for $\omega=0.798$, not only unbound but also superluminal orbits can be found. These two types of orbits belong to the region $r_{mb}^{in}<r<r_{mb}^{out}$ (see Fig.~\ref{fig:ex3}). If the condition $l_{\rm K}^{-}(r)=l_0$ is satisfied inside this region, neither cusps nor centers will be realized. The relation between $l_0$ and the torus topology can be found in Table~\ref{tab:types}.  

In contrast to the BS in the previous subsection, this type of BS will not be able to house Type 2$\bullet$ tori, for all double-center solutions will present a static surface. For example, by setting $l_0=-4.5M$, a Type 2$\circ$ torus can be found. The potential for such a torus, as well as the equipotential surfaces and the position of the static surface are depicted in Figs.~\ref{fig:pot_om_0671_lk45} to \ref{fig:epot_om_0671_lk45_2}.
We note that the center of the innermost torus in this case is inside the static surface and thus in prograde motion.

The innermost torus center is located at $r^{(1)}_{center}=0.66M$, near the ICO at $r_{\rm ICO}=0.51M$. The second center is at $r^{(2)}_{center}=11.8M$. For a polytropic fluid with $\Gamma=4/3$, this double-center torus has a density ratio of $\rho^{(1)}/\rho^{(2)}=8 \cdot 10^5$, thus the innermost center has a considerably higher density than the outermost. The cusp is found at $r_{\rm cusp}=5.26M$.

\begin{figure}

\begin{subfigure}{.5\textwidth}
  \centering
  \includegraphics[width=1\linewidth]{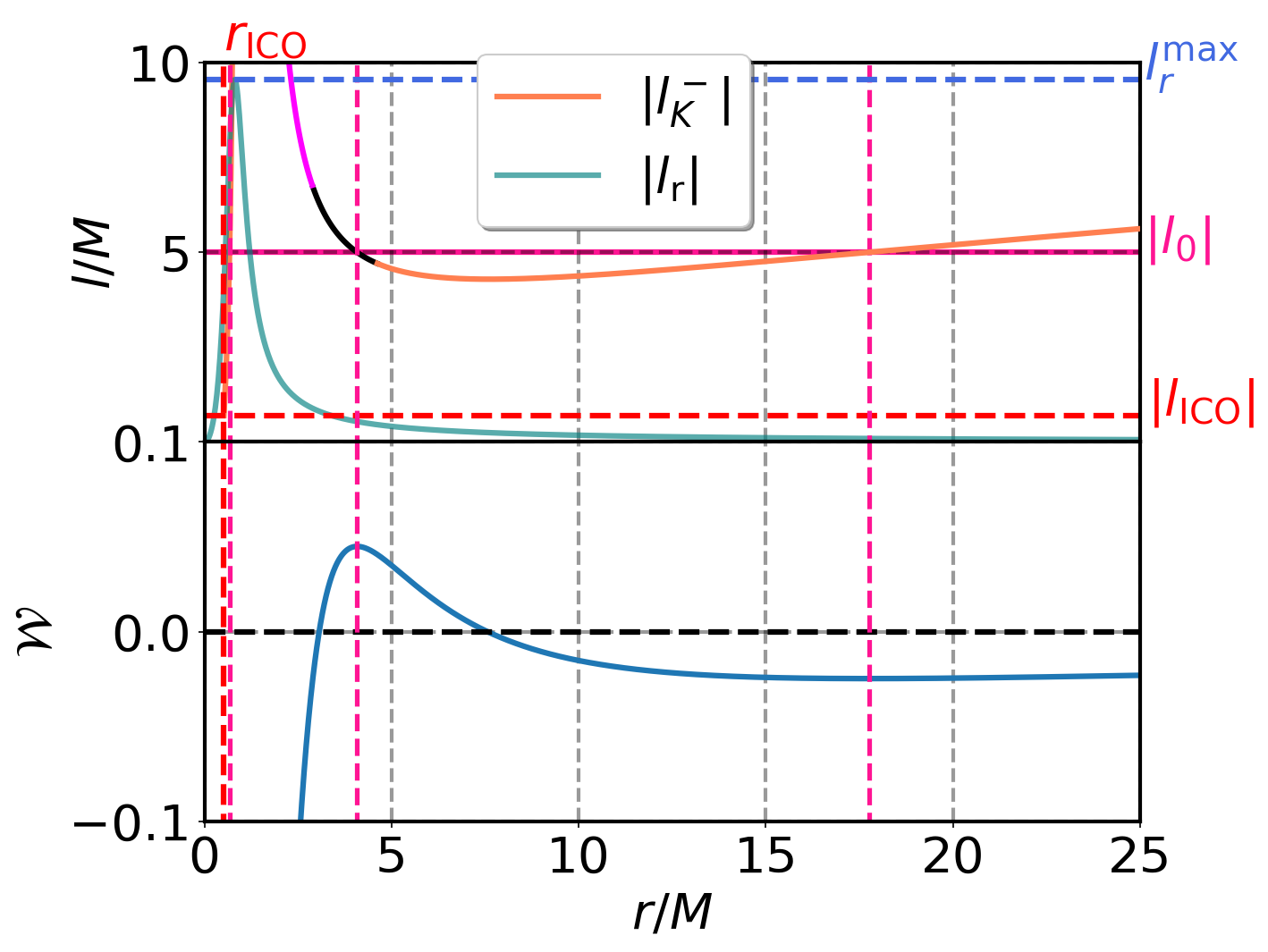}
  \caption{ }
   \label{fig:pot_om_0671_lk5}
\end{subfigure}
\begin{subfigure}{.5\textwidth}
  \centering
  \includegraphics[width=1\linewidth]{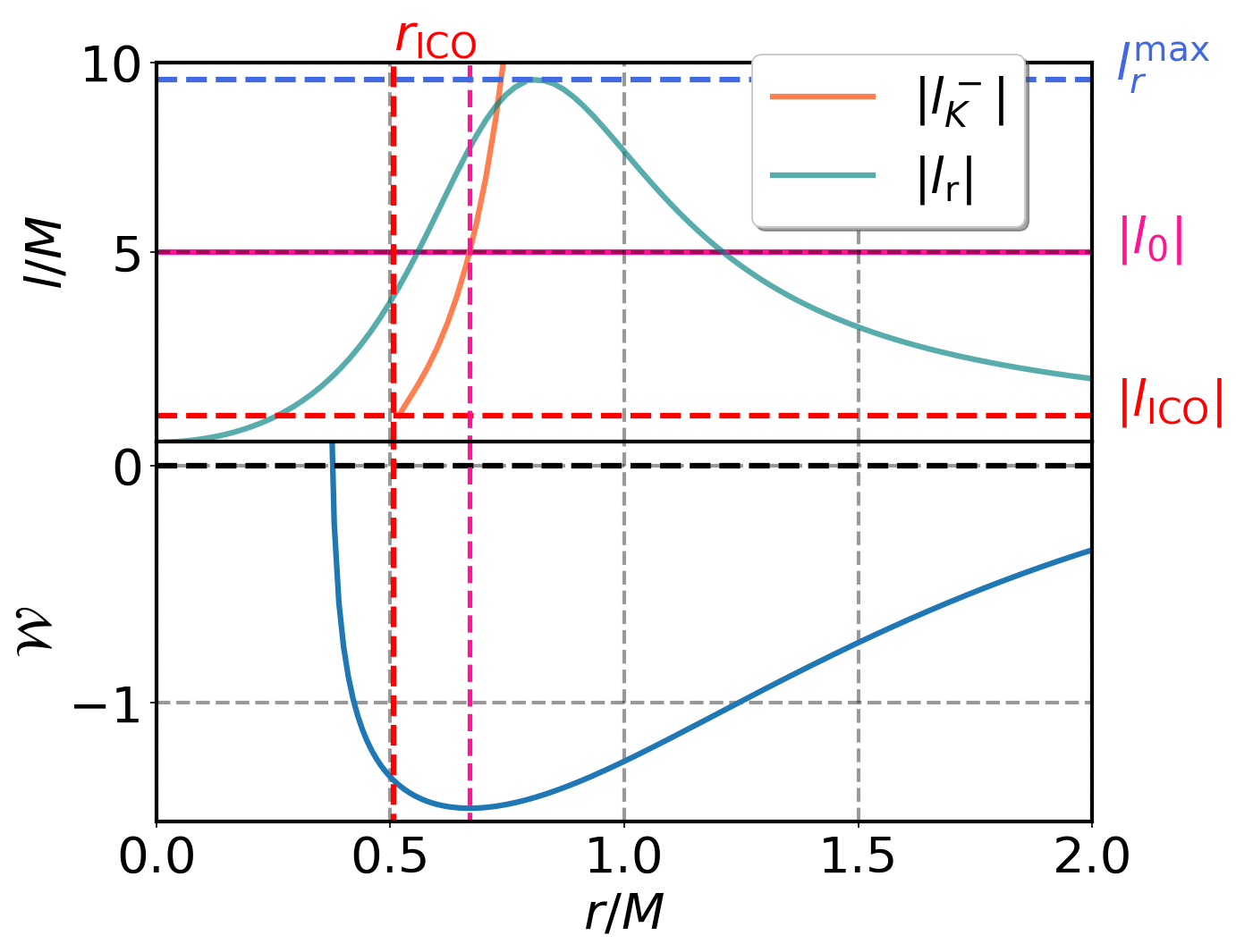}
  \caption{ }
   \label{fig:pot_om_0671_lk5_2}
\end{subfigure}

\begin{subfigure}{.55\textwidth}
  \centering
  \includegraphics[width=1\linewidth]{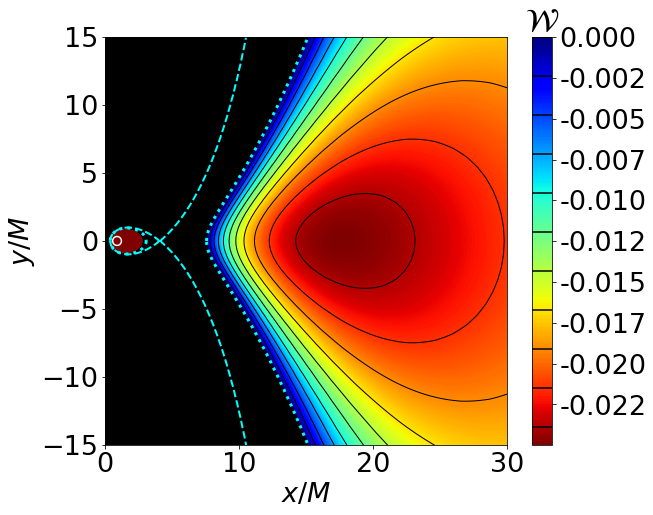}
  \caption{ }
  \label{fig:epot_om_0671_lk5}

\end{subfigure}%
\begin{subfigure}{.55\textwidth}
  \centering
  \includegraphics[width=1\linewidth]{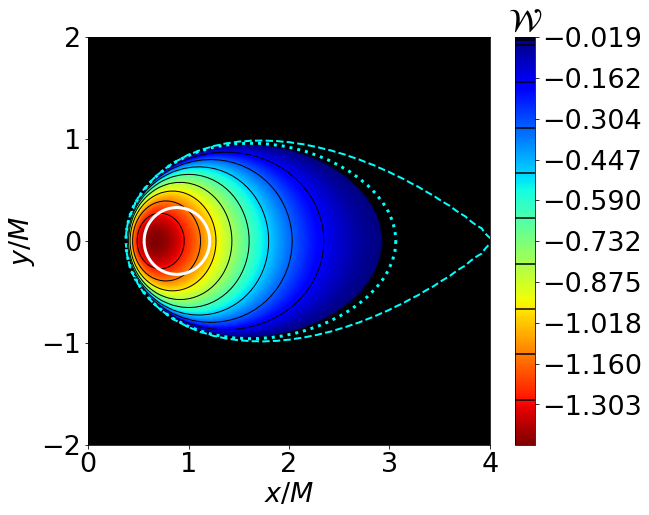}
 \caption{ }
   \label{fig:epot_om_0671_lk5_2}

\end{subfigure}\\

\caption{Example of a Type 3$\circ$ torus: (a - upper panel) Retrograde Keplerian specific angular momentum for a BS with $\omega=0.671$ and $k=1$. (a - lower panel) Potential for a torus with $l_0=-5M$ in the equatorial plane. The dashed magenta line marks the torus centers and position of the unfeasible cusp. (b) Same as (a) but with a different scale, highlighting the center of the innermost torus. (c) Equipotential surfaces for the same potential (meridional cross section). The dotted bright blue line represents the surface $\mathcal{W}(r,\theta)=0$, while the dashed bright blue line marks the self-intersecting equipotential related to the cusp. (d) Same as (c) but for a different range of $\mathcal{W}$ (see color maps) and with a different scale.}

\label{fig:eqp_om0671_l45}
\end{figure}

In order to exemplify the suppression of cusps in between the radii of the marginally bound orbits, we present a final example of a torus of Type 3$\circ$. By setting $l_0=-5M$, as seen in \ref{fig:pot_om_0671_lk5}, the condition $l_{\rm K}^{-}(r)=l_0$ is satisfied at three radial positions. For the innermost and outermost centers, located at $r^{(1)}_{center}=0.67M$ and $r^{(2)}_{center}=17.75M$, $\mathcal{W}^{(1)}$ and $\mathcal{W}^{(2)}$ assume negative values, therefore they can be included in closed torus solutions. In contrast, at the cusp position ($4.08M$) an unbound orbit can be found. Thus $\mathcal{W}(r_{cusp})>0$ and the cusp will sit on an open equipotential, depicted by the dashed blue line in Fig.~\ref{fig:epot_om_0671_lk5}. Belonging to an open equipotential, the cusp can not be achieved by a physical fluid configuration.  In fact, the largest possible outer border of the torus, $\mathcal{W}(r,\theta)=\mathcal{W}_{in}=0$, decomposes now into two disconnected surfaces. These surfaces are shown by a bright blue dotted line in Fig.~\ref{fig:epot_om_0671_lk5}. The torus is also endowed with a static surface containing the innermost torus center, as seen in Fig.~\ref{fig:epot_om_0671_lk5_2}. The ratio $\rho^{(1)}/\rho^{(2)}$ is $2\cdot10^6$.  

By increasing $l_0$ beyond $l_{mb}^{in}$, it is also possible, that the innermost center becomes also suppressed by being located at an unbound or superluminal orbit radius, yielding simply Type 1 tori for which the outermost center resides rather far from the BS center. 

As a final note we report that a static surface is also feasible for prograde tori, for $l_{\rm K}^{+}(r)$ can in fact assume negative values near the ICO. These surfaces ought to become more extended as $l_0$ increases and are fated to disappear as the whole torus achieves prograde motion (which will happen for $l_0>0$).  In fact, by also considering prograde tori, a further analysis on these surfaces' extension can me made. Given a BS, a torus endowed with a static surface ought to obey  $-l_{r}^{\max}<l_0<0$, with the threshold solutions $l_0=-l_{r}^{\max}$ and $l_0=0$ representing the ones housing the least and most extensive static surfaces feasible, respectively. In fact, the choice $l_0=-l_{r}^{\max}$ corresponds to a static surface degenerated into a ring while for $l_0=0$ the static surface ought to be closed at infinity. Hence, given a BS model in which a static surface is feasible, its extension can be arbitrarily small or large, depending only on the torus' specific angular momentum.

\section{Conclusion}

In this work we have explored the features of retrograde Polish Doughnuts around various BS solutions with winding number $k=1$. By setting the specific angular momentum distribution to be constant we were able to build simple versions of this class of disks. The tori were found to be endowed with a new type of structure, the static surfaces. These surfaces, a generalization of the static rings \cite{Collodel:2017end} for non-geodesic fluid motion, represent a surface in which the torus fluid would stay at rest with respect to a ZAMO at infinity. Inside these surfaces, matter flows in prograde motion, while outside it moves in a retrograde manner. Double-centered torus solutions were also realized, found before only for BSs with $k=4$ in the prograde case \cite{Meliani:2015zta}.  
  
In order to address these features, we have first analysed the specific angular momentum distributions of the Keplerian orbits for the BSs in question. The first novelty when considering retrograde orbits comes from the fact that some of these orbits were found to be unbound or even superluminal for some regions outside the ICO. Therefore new relations between the choice of the torus specific angular momentum and its topology had to be addressed. These relations now take into consideration the suppression of possible cusps or centers inside the annulus delimited by the innermost and outermost marginally bound orbits. Together with the rest specific angular momentum distribution, we were then able to classify the torus solutions, and track the conditions, both regarding $l_0$ and the BS solutions, for which each would be feasible. 

While showing examples of the torus types sheltered by our BS solutions, we were able to conclude the following:

Less relativistic BSs, represented by the solution with $\omega=0.960$, were shown to house tori with one center, no cusp and endowed with or without a static surface. For the case in which the tori featured a static surface, these fluid configurations were found to be rather thick and likely to be centered close to the ICO. Although possible, the range of solutions for which the torus center was placed inside the static surface, i.e. in prograde motion, was found to be minimal. 

Mildly relativistic BSs, represented by the solution with $\omega=0.798$, could in addition shelter also two-centered tori without a static surface. Further restrictions regarding the $l_0$ choice started to arise due to the Keplerian specific energy profile for these space-times. Meanwhile one-centered tori possessing a static surface were feasible for a wider range of $l_0$. Such surfaces were found to be more likely to contain the torus centers. 

Finally, the most relativistic BSs, represented by the solution with $\omega=0.671$, can house double-centered tori endowed with a static surface. In addition solutions of cusp-less double-centered tori are more abundant. Also, static surfaces become much more common around this type of BS, for the rest specific angular maximum becomes considerably larger than for the less relativistic stars.  

The study we have presented unveils some of the rich features that can be found for fluid configurations around BSs, by considering retrograde thick tori. Motivated by these results, further simulations on the evolution of this type of solutions as well as an investigation of the effects that self-gravity and magnetic fields could have on these structures could be an interesting topic future of research.

\section{Acknowledgements}

We gratefully acknowledge support by the DFG funded
Research Training Group 1620 ``Models of Gravity''.
LGC would like to acknowledge support via an
Emmy Noether Research Group funded by the DFG
under Grant No. DO 1771/1-1.
The authors would also like to acknowledge networking 
support by the COST Actions CA16104 and CA15117.

\section{References}
\providecommand{\newblock}{}

\appendix
\section{Field Equations}
\label{app:fe}

We list below the field equations (\ref{Einstein}) and (\ref{scaleq}) for the line element (\ref{metric}) in diagonal form.
\begin{align}
    r^2\partial_r^2\alpha+\partial_\theta^2\alpha &= -r\partial_r\alpha\frac{(2B+r\partial_rB)}{B}-\partial_\theta\alpha\frac{(2B\cot\theta+\partial_\theta B)}{B} \nonumber \\ &+\frac{B^2r^2\sin^2\theta}{2\alpha}\left[r^2\partial_r(\beta^\varphi)^2+\partial_\theta(\beta^\varphi)^2\right] \nonumber \\
    &+\frac{8\pi r^2 A^2f^2}{\alpha}\left[2(k\beta^\varphi-\omega)^2-m^2\alpha^2\right].
\end{align}

\begin{align}
    r^2\partial_r^2A+\partial_\theta^2A &= \frac{r^2(\partial_rA)^2+(\partial_\theta A)^2}{A}-r\partial_rA + \frac{AB^2r^2\sin^2\theta}{4\alpha}\left[r^2\partial_r(\beta^\varphi)^2+\partial_\theta(\beta^\varphi)^2\right] \nonumber \\
    &+ \frac{A}{\alpha B}\left[r\partial_r\alpha(B+r\partial_rB)+\partial_\theta\alpha(B\cot\theta+\partial_\theta B)\right]-8\pi A\left[r^2(\partial_rf)^2+(\partial_\theta f)^2\right] \nonumber \\
    &+\frac{8\pi A^3f^2}{\alpha^2B^2\sin^2\theta}\left[k^2\alpha^2-B^2r^2\sin^2\theta(k\beta^\varphi-\omega)^2\right].
\end{align}

\begin{align}
    r^2\partial_r^2B+\partial_\theta^2B &= -\frac{1}{\alpha}\left[r\partial_r\alpha(B+r\partial_rB)+\partial_\theta\alpha(B\cot\theta+\partial_\theta B)\right] - 3r\partial_rB-2\partial_\theta B\cot\theta \nonumber \\
    &+ \frac{B^3r^2\sin^2\theta}{2\alpha^2}\left[r^2\partial_r(\beta^\varphi)^2+\partial_\theta(\beta^\varphi)^2\right] - 8\pi A^2f^2\left(\frac{2k^2}{Br^2\sin^2\theta}+m^2B\right).
\end{align}

\begin{align}
    r^2\partial_r^2\beta^\varphi+\partial_\theta^2\beta^\varphi &=
    \frac{1}{\alpha B}
    \left[
    r\partial_r\beta^\varphi(rB\partial_r\alpha-4\alpha B-3\alpha r \partial_rB)+\partial_\theta \beta^\varphi(B\partial_\theta\alpha-3\alpha B\cot\theta-3\alpha\partial_\theta B)
    \right] \nonumber \\
    &+ \frac{32\pi k A^2f^2}{B^2\sin^2\theta}(k\beta^\varphi - \omega).
\end{align}

\begin{align}
    r^2\partial_r^2f+\partial_\theta^2f &=
    \frac{1}{\alpha B}
    \left[ 
    r\partial_rf(rB\partial_r\alpha-2\alpha B- \alpha r \partial_r B)-\partial_\theta f(B\partial_\theta\alpha+\alpha B\cot\theta+\alpha\partial_\theta B)
    \right] \nonumber \\
    &+ A^2f\left[ m^2r^2+\frac{k^2}{B^2\sin^2\theta}+\frac{r^2}{\alpha^2}(k\beta^\varphi - \omega)\right].
\end{align}

\end{document}